\documentclass[a4paper,fleqn]{cas-sc}

\usepackage[sort&compress, numbers]{natbib}
\usepackage[nodots]{numcompress}
\graphicspath{{figures/}} 
\usepackage{siunitx}
\DeclareSIUnit\watthour{Wh}

\usepackage{forest}

\usepackage{chngcntr}
 \counterwithin{figure}{section}
 \counterwithin{table}{section}
 \counterwithin{equation}{section}

\usepackage{acronym}

\def\tsc#1{\csdef{#1}{\textsc{\lowercase{#1}}\xspace}}
\tsc{WGM}
\tsc{QE}

\begin{document}
\let\WriteBookmarks\relax
\def\floatpagepagefraction{1}
\def\textpagefraction{.001}

\shorttitle{Comparison of different methods of spatial disaggregation of electricity generation}    
\shortauthors{O.\ Ravent{\'{o}s} et~al.\ }  


\title[mode = title]{Comparison of different methods of spatial disaggregation of electricity generation and consumption time series}

\tnotemark[1]
\tnotetext[1]{This research was funded as part of the \textit{MODEX-Net project} by the German Federal Ministry for Economic Affairs and Energy (BMWi) in the 6th energy research funding program. Funding Code: 03ET4074.}

\author[1]{Oriol Ravent{\'{o}}s}[orcid=0000-0002-0512-4331]
\cormark[1]
\ead{oriol.raventosmorera@dlr.de}
\credit{Conceptualization, methodology, formal analysis, investigation, data curation, writing---original draft preparation, writing---review and editing, supervision, project administration}
\author[2]{Thomas Dengiz}
\ead{thomas.dengiz@kit.edu}
\credit{Conceptualization, methodology, writing---original draft preparation, writing---review and editing, project administration}
\author[1]{Wided Medjroubi}[orcid=0000-0002-2274-4209]
\ead{Wided.Medjroubi@dlr.de}
\credit{Conceptualization, resources, writing---original draft preparation, writing---review and editing, supervision, project administration, funding acquisition}
\author[1]{Chinonso Unaichi}
\ead{Chinonso.Unaichi@dlr.de}
\credit{Conceptualization, methodology, formal analysis, investigation, data curation, writing---original draft preparation, writing---review and editing}
\author[3]{Andreas Bruckmeier}
\ead{ABruckmeier@ffe.de}
\credit{Conceptualization, data curation,  methodology, writing---original draft preparation, writing---review and editing, project administration}
\author[2]{Rafael Finck}
\ead{rafael.finck@kit.edu}
\credit{Data curation, writing---original draft preparation, writing---review and editing}
\affiliation[1]{organization={German Aerospace Center (DLR), Institute of Networked Energy Systems},
            addressline={Carl-von-Ossietzky Str.\ 15}, 
            city={Oldenburg},
            postcode={26129}, 
            country={Germany}}

\affiliation[2]{organization={Karlsruhe Institute of Technology (KIT), Institute for Industrial Production (IIP)},
            city={Karlsruhe},
            postcode={76187}, 
            country={Germany}}

\affiliation[3]{organization={Forschungsstelle f{\"{u}}r Energiewirtschaft (FfE) e.V.\ \slash Research Center for Energy Economics},
            city={Munich},
            postcode={80995}, 
            country={Germany}}


\cortext[1]{Corresponding author}

\begin{abstract}
Energy system models involve various input data sets representing the generation, consumption and transport infrastructure of electricity. Especially energy system models with a focus on the transmission grid require time series of electricity feed-in and consumption in a high spatial resolution. In general, there are two approaches to obtain regionalized time series: top\nobreakdash -down and bottom\nobreakdash -up. In many cases, both methodologies may be combined to aggregate or disaggregate input data. Furthermore, there exist various approaches to assign regionalized feed\nobreakdash-in of renewable energy sources and electrical load to the model's grid connection points. The variety in the regionalization process leads to significant differences on a regional scope, even if global values are the same.

We develop a methodology to compare regionalization techniques of input data for photovoltaics, wind and electrical load between various models as well as data assignment techniques to the power grid nodes. We further define two invariants to evaluate the outcome of the regionalization process at the NUTS 3 level, one invariant for the annual profiles and one for the installed capacities. This methodology enabled us to compare different regionalization and assignment workflows using simple parameters, without explicit knowledge of grid topology. Our results show that the resolution of the input data and the use of a top-down or a bottom-up approach are the most determinant factors in the regionalization process. 
\end{abstract}


\begin{keywords}
Spatial (dis)aggregation\sep 
(Dis)aggregation techniques\sep Regionalization\sep 
Electricity generation\sep 
Electricity consumption\sep 
Load time series\sep 
Energy systems\sep 
Aggregation techniques comparison\sep 
Power system models\sep 
Model comparison
\end{keywords}

\maketitle

\section{Introduction}

\label{sec_intro}


Energy system modeling has seen a large growth in the last two decades largely due to the challenges of increasing greenhouse gas emissions and global climate change. \acfp{ESM} aim at modeling current as well as future energy system configurations using simulation and/or optimization techniques in order to explore, investigate and evaluate different possible scenarios. Different types of \acp{ESM} exist based on the particular modeling approach, scope, methodology and the level of spatial and temporal resolutions considered \cite{bhattacharyya2010review, keirstead2012review, nakata2004energy}. Energy sector stakeholders such as policy makers, system operators, utilities and investors rely on the results of \acp{ESM} to formulate energy-related goals, economic and policy decisions, as well as to plan and operate the energy system \cite{cao2020modeling, TransnetBWStromnetz2050Studie, xiao2020modeling}. Because of their high relevance and impact on current and future energy policies as well as operating and planning decisions, it is important to interpret and assess outcomes of various \acp{ESM} and understand their divergence. 

\begin{figure}
\centering
\fcolorbox{black}{white}{
\begin{minipage}{0.6\textwidth}
{\footnotesize
\noindent {\bf{List of abbreviations}}
\begin{acronym}[ENTSO-E]
\acro{RES}[RES]{Renewable Energy Source}
\acro{BU}[BU]{Bottom-Up}
\acro{TD}[TD]{Top-Down}
\acro{PV}[PV]{Photovoltaics}
\acro{EEG}[EEG]{Erneuerbare-Energien-Gesetz}
\acro{MaStR}[MaStR]{Marktstammdatenregister}
\acro{OPSD}[OPSD]{Open Power System Data}
\acro{ENTSO-E}[ENTSO-E]{European Network of Transmission System Operators for Electricity}
\acro{RMSE}[RMSE]{Root Mean Square Error}
\acro{ehv-nodes}[ehv-nodes]{extra high voltage nodes}
\acro{NUTS}[NUTS]{Nomenclature of territorial units for statistics}
\acro{LAU}[LAU]{Local Administrative Unit}
\acro{ESM}[ESM]{Energy system model}
\end{acronym}
}
\end{minipage}
}
\end{figure}

Comparing different \acp{ESM} is a fruitful way to improve and verify existing models. There have been a number of activities pursuing this goal. For instance, at the international level, there is the Energy Modeling Forum created in \num{1976} \cite{huntington1982modeling} and, at the EU level, there are the Energy Modelling Platform for Europe (EMP-E) \cite{muller2018comprehensive}, the ACROPOLIS project (Assessing Climate Response Options: POLIcy Simulations - Insights from using national and international models) \cite{das2007acropolis}, and the project CASCADE-MINTS (CAse Study Comparisons And Development of Energy Models for INtegrated Technology Systems) \cite{mints2003case}. In addition, there are ongoing projects like RegMex \cite{oei_4537}, which focuses on scenarios and sector coupling comparison, and the 4NEMO project \cite{savvidis2019gap}, which focuses on socioeconomic modeling and the capabilities of models to answer policy-related questions. To our knowledge, however, there is no project that compares the spatial aggregation and disaggregation processes of transmission grid models (beyond comparing just the final spatial resolution). These processes include the assignment of electricity feed-in and load to the \acfi{ehv-nodes}, also referred as \emph{grid nodes} in this article for simplicity. These spatial aggregation and disaggregation processes will be referred to as \emph{regionalization}. The main difficulty to carry out such a comparison is the insufficient information provided on the model's inner workflows as well as the large amount of input and output data sets involved in the regionalization. Graphic representation of regionalization workflows also differ widely and are difficult to compare, e.g. \cite[Fig.~1]{Boeing2017}, \cite[Figs.~6.4 and 6.14]{EsserFrey2012_1000028367}, \cite[Abb.~3]{Koch2015}, \cite[Fig.~4]{priesmann2021time},   \cite[Fig.~2]{Slednev2017regionalizing}, and \cite[Fig.~2]{slednev2018highly}.

The present paper introduces a method for comparing regionalization workflows from input data to the power grid nodes. The developed method is then applied to eight German transmission grid models using a 2016 scenario: ELMOD, eTraGo, Europower, ISAaR, MarS\slash ZKNOT, MILES, PERSEUS and PowerFlex, which are members of the MODEX-Net project (BMWi 03ET4074) \cite{modex_net2021}. There is a short description of the models in \cite[Appendix]{modex_AP1}. In the framework of MODEX-Net project, an overall comparison of the models under study is already provided in \cite{modex_AP1}, whereas a comparison of the optimization of fixed scenarios is presented in \cite{modex_AP3}.

This paper is organized as follows: Section \ref{sec_def_lit_rev} contains a literature review and introduces basic concepts related to regionalization. The methodology used for the models comparison is introduced in Section \ref{sec_data_method}. The results of the comparison method applied to the eight different German transmission grid models in MODEX-Net are presented in Section \ref{sec_results}, followed by a discussion in Section \ref{sec_discussion}. Finally, in Section \ref{sec_conclusion} we lay out the main conclusions of the article. Two appendices are included. Appendix \ref{sec_NUTS3plots} contains the plots of the regionalization of the installed capacities and annual electricity demand of the models under study and Appendix \ref{sec_NUTS2plots} recomputes the invariants given in Section \ref{sec_results} using a broader spatial resolution, in order to highlight the relevance of comparing regionalized data at the finest level available.

\section{Basic concepts and literature review}
\label{sec_def_lit_rev}

We define regionalization as the process to allocate input data to a transmission grid node in an \acs{ESM}. There are two main regionalization approaches: \acfi{TD} and \acfi{BU}. Before providing a concrete definition and giving examples of both approaches, it has to be stated that both names have a different meanings when applied to models as a whole (not only to the regionalization). For instance, in \cite{amerighi2010models, connolly2010review, grubb1993costs, hall2016review, herbst2012introduction, muller2018comprehensive, pfenninger2014energy, prina2020classification}, a Top-Down model is synonym of macroeconomic model, whereas a Bottom-Up model is one that includes technical components with their individual characteristics and is used for a technical analysis via an optimization process.

A \acs{TD} regionalization process uses input data of a large area (like a whole nation) generally obtained from an energy scenario, to distribute it to smaller regions. An example of a \acs{TD} regionalization would be the disaggregation of future electricity demand for charging electric vehicles to a lower administrative level, given a certain number of vehicles and annual demand for charging at country level obtained from a national scenario. It is possible to use either the current regionalized electricity demand as a regional factor or a distribution key to distribute the given global energy demand to the regionalized electric vehicle demand. Alternatively, regional statistics of registered cars as distribution key can be used.

A \acs{BU} \emph{regionalization} process uses input data from low scale regions (or even single geographical locations) to aggregate it to a larger region. Considering the example above of future electricity demand for charging electric vehicles, one \acs{BU} regionalization could be as follows: Based on a diffusion model and some indicators like wealth, the number, type and usage of electric vehicles is determined for each region. Finally, the aggregated individual profiles determine the regionalized demand. 

An \emph{hybrid} regionalization is a combination of both \acs{TD} and \acs{BU}. For instance, in \cite{priesmann2021time}, the demand regionalization is done separately for four sectors (residential, industrial, commerce, and mobility) by combining the annual consumption at a municipality level with load profiles at national level to produce a final regionalized load profiles at an intermediate level.

Clustering methods can also be considered as \acs{BU} regionalization since they aggregate regions together to reduce the complexity of the model. In \cite{martinez2021review}, two types of clustering methods are investigated: The non-optimal clustering methods which are based on geographical regions, such as a country, a district or a state and the optimal clustering methods which are based on spatial clustering methods, such as k-\nobreakdash means or max-\nobreakdash p regions. Although optimal clustering techniques are widely used in energy system modeling and an extensive bibliography exists addressing the trade-\nobreakdash off between the number of clustering regions and the computational accuracy, cf.\ \cite{kueppers2020data, frysztacki2020modeling, frysztacki2021, klein2016aggregation, horsch2017role}, optimal clustering is usually not used in the \acs{BU} regionalization process for transmission grid models, as it would deform the underlying transmission grid. 


We will compare the regionalization processes at \acs{ehv-nodes} level. This is problematic for two reasons. First, even if two models use the same reference network (or power grid), the topology can still be different. Hence, there is a need to compare the data aggregated to some common resolution. The other problem is that most \acp{ESM} do not make public the data at the grid node level. For instance, in order to validate annual consumption, in \cite{priesmann2021time}, the authors have to aggregate the profiles at national level. To give another example, in the RegMex project  \cite{gils2019comparison}, a comparison between four models of the German power grid is done, but before the comparison, all the models are clustered to \num{18} regions and the input data is harmonized, thus, diluting the regional differences. In the present article, we conduct the comparison of the output of the regionalization process at NUTS 3 level, where \acs{NUTS} stands for \emph{Nomenclature des unit\'{e}s territoriales statistiques}, which is a geocode standard for subdividing countries for statistical purposes developed in the European Union. In the case of Germany, NUTS 0 is just Germany as one region, NUTS 1 are the federal states, NUTS 2 are government regions, and NUTS 3 are districts. There is a finer division to municipalities called \acfi{LAU}.


Grid connection points of conventional and large scale \acfi{RES} plants are normally explicitly known, as they are either connected to \acs{ehv-nodes} or to \SI{110}{\kilo\volt} nodes. For decentralized electricity demand and \acs{RES} feed-in, grid connection points are in the lower voltage distribution grid. Since in most \acp{ESM} the distribution grid is not considered, different methods emerged to assign electricity feed-in and demand to \acs{ehv-nodes}. Most of them use the so-called \emph{grid assignment regions}. These are regions around the \acs{ehv-nodes} used to assign generation and load data located in this region to the ehv-\nobreakdash node. Another type of regions used in regionalization are the regions where the data is available, i.e \emph{data regions}, e.g.\ postal code areas, municipalities or districts. The use of node and data regions for assigning input data to a grid node are at the core of the regionalization process, but they are mostly absent in most of the \acs{ESM} literature reviews.


Next, we review the literature for \acp{ESM} comparison and regionalization processes. Most papers focused on regionalization, just compare different spatial resolution of data for the same model, usually evaluating clustering algorithms in terms of the trade-off between optimization error and computational time. However, there is a gap in the literature concerning the comparison of structural regionalization processes.

There exists an vast literature comparing \acp{ESM} going back to \cite{greenberger1979assessing}, which identifies two broad strategies for comparing models: Compare the structure of the models or compare the specific application of the models. Our study falls in the first category. If we were to compare the models discussed in this article regarding their application scope, they would all fall into a similar category (i.e. technical optimization models for Germany in the European context). 

A succinct literature review of energy model comparisons is provided in \cite{muller2018comprehensive}, where it is made clear that in order to characterize or classify current \acp{ESM}, a long list of dimensions is needed. For instance, in \cite{hiremath2007decentralized} (following the work of \cite{grubb1993costs, jebaraj2006review, hourcade1996estimating}), the authors suggest a classification with the following \num{9} dimensions: General and specific purposes; model structure; analytical approach; underlying methodology; mathematical approach; geographical coverage; sectoral coverage; time horizon; and data requirements. Although \cite{hiremath2007decentralized} focuses on decentralized energy planning, the processes of \emph{decentralization} are not compared and only the final resolution level of the decentralization is reported (either village, block, district or state level). Even if the regionalization process of assigning installed capacities, availability and demand profiles at the level of \acs{ehv-nodes} can differ widely, its precise characterization is missing in most energy model review articles, e.g.\ \cite{amerighi2010models, connolly2010review, herbst2012introduction, hourcade2006hybrid}. In \cite[pag.~8]{aryanpur2021review} and \cite{martinez2021review}, it is pointed out that there are two main aspects that limit the spacial resolution of \acp{ESM}. One is the available data and the other is the lack of transparency of the regionalization process. In \cite{aryanpur2021review}, \num{36} national-scale energy system optimization models are reviewed and it is concluded that the low level of disaggregation can both under- and overestimate some of the costs resulting in possibly misleading outcomes of the optimization. However, just \num{4} of the models analyzed in \cite{aryanpur2021review} consider hourly resolution data and only the final number of regions is reported \cite[Table~1]{aryanpur2021review}. In \cite{martinez2021review}, \num{34} \acp{ESM} are analyzed and the need of high resolution \acp{ESM} is highlighted. In \cite{siala2019impact}, an example of the optimization of an European energy system is provided for 2015 and 2050. The aggregation is made by successively using the k-means clustering and the max-p regions method to obtain spatially connected clusters. The authors found that most \acp{ESM} use administrative regions based on data availability and propose an alternative method to define aggregation regions based on weather data. Similarly, in \cite{fleischer2020minimising}, the max-p regions clustering algorithm is used to derive aggregation regions smaller than nationalities that diminish the two negative effects of a broad \acs{ESM} resolution, namely, increase in solar and wind capacity and decrease in transmission capacity. In \cite{martinez2021review, fleischer2020minimising, siala2019impact} the authors motivate the use of non-administrative regions, but they refer to federal or national regions. This becomes less relevant if the administrative regions are small (like the NUTS~3 regions used in the present article). In \cite[Table~1]{savvidis2019gap} a classification of energy model reviews is provided, based on four non-excluding focus categories: Model description; classification scheme; field of use; and identification of suitable models. The present paper falls in the category of papers providing a classification scheme, but focused just on the rationalization process. In \cite[Figure~1]{savvidis2019gap}, they propose a classification of \acp{ESM} based on \num{30} criteria, following the work of \cite{hall2016review}, but also using input from \cite{beeck1999classification, bhattacharyya2010review, cao2016raising, connolly2010review, pfenninger2014energy}. As in the case of \cite{muller2018comprehensive}, none of these criteria allows for a differentiation of the regionalization processes as we intend in the present article. Again in the case of \cite{gils2019comparison} all the models considered in the current paper would fall in the category of ``\acs{BU} engineering models''. In \cite{gholami2020upscaling}, a classification of energy forecasting models is proposed dividing them into 13 categories without addressing the influence of the regionalization process.

In \cite{simoes2017impact}, Simoes et al. assess the impact of the spatial aggregation on the deployment of PV and wind by an \acs{ESM} for the years 2030 and 2050. They use the JRC-EU-TIMES model for Austria comparing the electricity generation differences when considering 2 or 79 wind regions and 1 or 5 PV regions. They show that disaggregating wind regions results in lower electricity generation from wind and PV (up to 80\% less of wind generation). In the present paper, we do not compare a model with itself using different regionalization resolutions, but different models at the finest regionalization resolution possible. The impact of spatial aggregation is also investigated in \cite{frew2016temporal}, where the authors highlight the effect of the spatial resolution on wind and solar site development. This was done by investigating individual site build-out versus uniform development across all sites. The spatial resolution, which is related to the spatial aggregation method, had large impacts on total system costs. The results reveal the importance of taking into account the spatial resolution of each step of the regionalization process. 


The above mentioned papers do not address the challenging issue of comparing the regionalization process and the data aggregation at the grid node level. The novelty of the present contribution lies in the accurate representation of the (German) transmission grid by all the investigated models. The aim behind our method is to provide an approach to identify the main characteristics of the different regionalization processes, from input data to grid nodes and compare its output at the local level. 

\section{Methodology}
\label{sec_data_method}



The methods and results presented in the paper are obtained from the models which are part of the the MODEX\nobreakdash-Net project \cite{modex_net2021}: ELMOD, eTraGo, Europower, ISAaR, MarS\slash ZKNOT, MILES, PERSEUS and PowerFlex. Since three manuscripts were produced in this project (the other two being \cite{modex_AP1} and \cite{modex_AP3}), a brief description of the models is only included in \cite[Appendix]{modex_AP1}. The project MODEX\nobreakdash-Net focuses on the comparison of transmission grid models of Germany (\SI{220}{\kilo\volt} or higher) and its neighboring countries.

As stated in Section \ref{sec_def_lit_rev}, the process of allocating input data into the grid model components is called regionalization. In \cite{Slednev2017regionalizing}, four categories of input data are considered for \acp{ESM}: transmission grid data (nodes and lines), conventional power plants data (location and capacity), renewable energy resources (\acs{RES}) data (spatial resolution and generation profiles) and demand data (spatial resolution and demand profiles). The regionalization process transforms this input data into \emph{grid components} of the output model. These grid components are allocated to \acs{ehv-nodes}. In this paper, we focus on the following grid components: \acfi{PV}, wind energy (onshore and offshore) and electricity demand. For these grid components we will analyze installed capacities, generation profiles and demand profiles. In Section \ref{sec_reg_process} a method to compare regionalization workflow is described and in Section \ref{sec_indicators} two indicators to compare its output at a local level are introduced.

\subsection{Regionalization process comparison}
\label{sec_reg_process}

In this section we provide a method to compare regionalization workflows. We focus our analysis on three aspects: capacity or demand scalar factors (from now on referred to as \emph{factor}), temporal hourly profiles (form now on referred to as \emph{profile}) and post-processing corrections. For instance, in the case of \acsp{RES}, the capacity factor can be the installed capacity and the profiles can either come from weather data, historic feed-in profiles or from standard load profiles.

In order to assign data to a node, most models use \emph{grid assignment regions}. A model can use either \emph{node regions} and assign all the data of a region around a certain node to this unique node (\acs{BU}) or \emph{data regions} distributing the data to all nodes in the data region (\acs{TD}). The spatial resolution of this region depends on the grid resolution itself, which might not be homogeneous throughout the considered territory. Nevertheless, the different ways to define the grid assignment regions can have a huge influence on the regionalization process and its output. A brief description of the regionalization process used in each model can be found in the following list, highlighting its resolution in terms of NUTS level when possible: 

\begin{itemize}
    \item ELMOD: for \acs{RES}, node regions defined by Voronoi cells of the \acs{ehv-nodes} are used, but the load is distributed evenly to the \acs{ehv-nodes} inside each NUTS 3 region.
    \item eTraGo: node regions starting from a \acs{LAU} region are assigned to the \SI{110}{\kilo\volt} node belonging to it. If two nodes are in one region, it is subdivided using Voronoi cells of the nodes belonging to it. If there is no node in one \acs{LAU} region, values get assigned to the closest node. Finally, each \SI{110}{\kilo\volt} node regions is assigned to an ehv-node region using the shortest grid-path (\emph{routing}) \cite{huelk2017}.
    \item Europower: node regions are defined by Voronoi cells of the \acs{ehv-nodes}.
    \item ISAaR: Data regions are used by assigning data in NUTS 3 areas to the grid node contained in it. If there is no node in the NUTS 3 region, it is assigned to the next-neighbour grid node. If there are more grid nodes within one NUTS 3 area, the assignment is conducted proportionally by area.
    \item MarS\slash ZKNOT: use node regions starting from a \acs{LAU} region, then assigning to \SI{110}{\kilo\volt} nodes using weighting distance and then to an ehv\nobreakdash-node region using the shortest grid-path.
    \item MILES: data regions are defined by first assigning load and \acs{RES} data to the centroid of a \acs{LAU} region and then distributed to the \acs{ehv-nodes} within a threshold radius proportionally based on their distance to the centroid. 
    \item PERSEUS: data regions are assigned to \SI{110}{\kilo\volt} nodes by area weights from the interception of the data region (NUTS 3 for demand) and Voronoi cells around the  \SI{110}{\kilo\volt} nodes. The  \SI{110}{\kilo\volt} nodes is assigned to an ehv-node using the shortest grid-path (\emph{routing}).
    \item Powerflex: node regions are defined by Voronoi cells around the \acs{ehv-nodes}.
\end{itemize}

In order to classify the types of regionalization processes of factors and profiles, we identified four paths for assigning data to a grid node which are listed in the following (notice that, the data can be associated to a region or to a concrete geographical location):

\begin{itemize}
    \item Assign local data to the closest node.
    \item Assign local data in a node region to the unique node present in this region.
    \item Distribute the data in a data region to the nodes located in this region.
    \item Overlap of data regions with node regions.
\end{itemize}

From the regionalization approaches cited above, we highlight the spatial resolution of the data in terms of NUTS level, the source of the data and, in a broad sense, whether they are \acs{BU} or \acs{TD} approaches. In the case of the profiles, we also highlight whether they are created from weather data or based on standard load profiles. For the post-processing correction after the node assignment, we highlight the spatial and temporal resolution of this correction.

In the current paper, we only consider a historical scenario, but in the case of future scenarios the regionalization of the future installed capacities should be added to the analysis. Concretely, one should list the local parameters that are taken into account for the future potential of a region and differentiate between the models that use an extension planning optimization problem to determine the future installed capacities, e.g.\ \cite{horsch2018pypsa, Slednev2017regionalizing}, and the ones that take the values from a given scenario, cf.\ \cite{muller2014impact}.

\subsection{Comparison of the regionalization output}
\label{sec_indicators}

In this section we describe two invariants to compare the output of a regionalization at a local level. As pointed out in Section \ref{sec_reg_process}, we focus on two types of regionalization data output: factors and profiles. In the case of \acs{RES}, the factor of interest is the installed capacity factor. Since PowerFlex and Europower do not use installed capacities, we derived them from the profiles, by means of the full load hours in Germany from the reference \cite{entsoe_factsheets2016}. The full load hours we considered are \SI{875}{\hour} for \acs{PV}, \SI{1465}{\hour} for wind onshore and \SI{2927}{\hour} for wind offshore. The installed capacity is then computed for each generator as the sum of its power profile divided by the full load hours. In case of the demand, the factor of interest is the annual demand. If we were to study a future scenario, we would have to use potential capacity factors.

For the comparison of profiles, it is important that the models use the same weather year with hourly profiles ranging from 01.01.2016 to 30.12.2016. However, Europower and ISAaR use the years 2015 and 2012; respectively for both \acs{RES} and load which could not be adapted to 2016. To be precise, ISAaR uses 2012 profiles but factors them so that the full load hours correspond to the ones of 2016. Europower uses 2015 data and the installed capacities are computed using full load hours as indicated above. For this reason, it is expected that Europwer and ISAar will have larger differences with the other models when comparing \acs{RES} profiles. In the case of the load profiles, though, we adapt the load profiles so that we can match the weekly pattern. Precisely, we shift the profile forward from January second and fill in the missing gap by the corresponding days from the end of the year. This ensures that in all models the year starts with a Friday. The load profiles of MILES are adapted in a similar way, since MILES uses load profiles from 2017. For eTraGo, it should also be mentioned that, although the 2016 calendar year is used, the ground data to calculate the load profiles (GDP, population and land use) is from 2011.

The main issue for comparing the regionalization output is that this data is allocated on the \acs{ehv-nodes} and each model uses a different grid, although all represent the German transmission grid. This is due to the different data sources to derive the grid data as well as the aggregation levels used in each model (see \cite{modex_AP1} for an overview of the transmission grid modeling and the number of nodes considered in each model). Furthermore, not all models can provide the output of the regionalization process in full detail due to licensing issues. For this reason, the output of the regionalization process will be aggregated at NUTS 3 level. This enables a high local resolution without needing the specific coordinates of the nodes. Notice that, even if the data was available at the nodal level, the analysis at NUTS 3 level proposed in this article is appropriate to make the different ehv\nobreakdash-grids comparable. On one hand, NUTS 3 regions are small enough to separate most ehv\nobreakdash-nodes and, on the other hand, NUTS 3 administrative borders are better related to the transmission grid topology than other abstract regions defined by the tessellation of the territory using squares or Voronoi cells. Nevertheless, our method is sensitive to the region definition, since the different models can allocate the same generator to nodes located in different regions. Therefore, it is important to also consider coarser region levels, like NUTS 2, to check if the differences or similarities between the models stabilize when considering larger regions. Even more challenging is the case of wind offshore, as some models might allocate it to an offshore region, which is the case for Europower. In principle, we could assign each offshore region to the NUTS 3 region connected to it. However, this is not possible for Europower, since the offshore capacity is distributed in the later optimization process and is not clearly allocated to one onshore region. In this case, we assign the capacity to the closest onshore region, but we also compare wind offshore data using only two regions: North Sea and Baltic Sea, to study the validity of our assignment.

In order to compare the installed capacities and the annual demand at NUTS 3 level, we compute the \acfi{RMSE} across all NUTS 3 regions for each pair of models, assigning zero to the regions with no installed capacity:
$$
\mbox{\rm RMSE}\left(\{x_t\}_{t=1\dots T}, \{y_t\}_{t=1\dots T}\right) = \sqrt{\frac{\sum_{t=1}^{T}\left(x_t - y_t\right)^2}{T}},
$$
where $\{x_t\}_{t=1\dots T}$ and $\{y_t\}_{t=1\dots T}$ are the installed capacities for each NUTS 3 region in Germany of the two models that are being compared. This approach results in \num{28} values comparing each pair of models, which will be plotted as a symmetric $8\times 8$ matrix with zeros on the diagonal. Notice that, the outcome of such statistics is no qualitative nor quantitative indicator of the regionalization, it rather indicates how similar this group of \num{8} models are. Even more, the fact that models do not have installed capacity in every region (since there are regions with no \acs{ehv-nodes}), can make the {\acs{RMSE}} misleading in absolute terms, as it divides by the total number of NUTS 3 regions in Germany, i.e.\ {\num{401}}. This is especially the case for wind offshore. Hence, \acs{RMSE} has only to be viewed in relative terms.

To compare the profiles, we compute, for each pair of models, the mean value of the \textit{Pearson correlation} of its profiles across all NUTS 3, where both models have some installed capacity. We use only the regions with some installed capacity so that the Pearson correlation is well defined. It is also important to mention that, we use absolute profiles with energy availability per region, in contrast to relative profiles with the percentage of energy available from the installed capacity. 

\section{Results}
\label{sec_results}

\subsection{Regionalization workflows classification}

In this section, the regionalization workflows for \acs{PV}, wind onshore, wind offshore and demand are classified according to Section \ref{sec_reg_process}. We study separately the regionalization of hourly profiles and factors (annual demand and installed capacity).

\subsubsection{Photovoltaics}\label{subsec_PV_results}

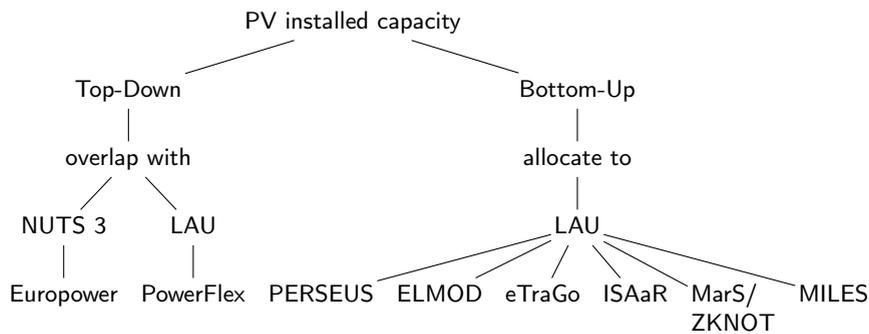
\begin{figure}[htbp]
\centering
{
\begin{forest}
for tree={align=left, inner sep=1pt, text=black, edge = {black}}
[PV installed capacity
    [Top-Down 
        [overlap with
            [NUTS 3
                [{Europower} ]]
            [LAU
                [{PowerFlex} ]]]]
    [Bottom-Up 
        [allocate to
            [LAU
                [{PERSEUS} ]
                [{ELMOD} ]
                [{eTraGo} ]
                [{ISAaR} ]
                [{MarS\slash\\ ZKNOT} ]
                [{MILES} ]]]]]
\end{forest}}
\caption{Dendrogram of regionalization of PV installed capacity workflows}
\label{forest_PV_installed}
\end{figure}

A summary of the regionalization workflow for the installed \acs{PV} capacities of all models can be found in Table \ref{table_PV_installed_cap}. The data sources of the different models are mainly the \acfi{EEG}, the \acfi{MaStR}, \acfi{ENTSO-E} (\cite{Entsoe2016_transparency}) or the \acfi{OPSD} (\cite{OPSD-project-website}). Two  different workflow groups are distinguished, the models using a \acs{TD} approach: Europower (at NUTS 3 level) and PowerFlex (using input at \acs{LAU} and NUTS 1 levels) and the rest of the models using a \acs{BU} approach, allocating each generator to the node region to which it belongs. These groups can be visualized as the dendrogram in Figure \ref{forest_PV_installed}. Notice that, MILES uses a different regionalization method, first by assigning all the generators in a \acs{LAU} region to its centroid (\acs{BU}) and then distributing the capacity to the buses inside a threshold radius. We consider the MILES method to be a \acs{BU} method since it essentially starts at the generator location and assigns it to the closest node.

\begin{table}[htbp]
\caption{Regionalization workflow characteristics of \acs{PV} installed capacity for each model.}
\label{table_PV_installed_cap}
\begin{tabular*}{\tblwidth}{@{}LL p{40mm} p{70mm}@{}}
 \toprule
 & \bf{\acs{BU}/ \acs{TD}} & \bf{Input resolution \& \newline source} & \bf{Node allocation} \\ \midrule
 ELMOD & \acs{BU} & \acs{LAU} Register (\acs{EEG}) & Generator belonging to a node region \\
 eTraGo & \acs{BU} & \acs{LAU} Register (\acs{OPSD}) & Generator belonging to a node region\\
 Europower & \acs{TD} & NUTS 2 Register (\acs{ENTSO-E}) \newline \& NUTS 3 distribution & Overlap of generation regions and node regions \\
 ISAaR & \acs{BU} & \acs{LAU} Register (\acs{MaStR}) & Generator belonging to a node region \\
 MarS\slash ZKNOT & \acs{BU} & \acs{LAU} Register (\acs{EEG}) & Generator belonging to a node region  \\ 
 MILES & \acs{BU} (\& \acs{TD}) & \acs{LAU}  Register (\acs{EEG}) & Generator belonging to a \acs{LAU} region\newline (\& distributed to nodes in the allocation region) \\
 PERSEUS & \acs{BU} & \acs{LAU} Register (\acs{EEG}) & Generator belonging to a node region \\
 PowerFlex & \acs{TD} & \acs{LAU} Register (\acs{EEG}) \& \newline NUTS 1 statistical data & Overlap of \acs{LAU} generation region and node regions  \\
 \bottomrule
\end{tabular*}
\end{table}

\begin{figure}[htbp]
\centering
{
\begin{forest}
for tree={align=left, inner sep=1pt, text=black, edge = {black}}
[PV profiles
    [Power profiles / Top-Down 
        [distributed from
            [NUTS 0
                [{ELMOD} ]
                [{PowerFlex} ]]]
        [overlaped with
            [NUTS 2
                [{Europower} ]]]]
    [Weather data / Bottom-Up 
        [of node region
            [NUTS 3
                [{ISAaR} ]]
            [LAU
                [{MarS\slash\\ ZKNOT} ]
                [{MILES} ]
                [{PERSEUS} ]]]
        [of closest point
            [LAU
                [{eTraGo} ]]]]]
\end{forest}}
\caption{Dendrogram of regionalization of PV profiles workflows.}
\label{forest_PV_profiles}
\end{figure}
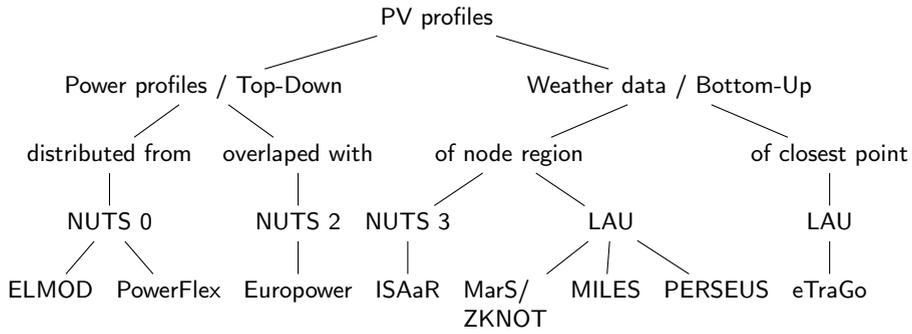

The \acs{PV} profile regionalization process of the different models is summarized in Table \ref{table_PV_profiles} and the corresponding dendrogram in Figure \ref{forest_PV_profiles}. We distinguish two different groups: ELMOD, Europower and PowerFlex are using profiles for regions and then distributing them into the nodes (\acs{TD}), although the data used by Europower is based on a \acs{BU} computation of potential power in NUTS 2 regions. The rest of the models use weather data to produce profiles (\acs{BU}). In this last group, eTraGo uses the closest weather data point to the node whereas the rest use all the points in the weather data grid belonging to the node region. 

\begin{table}[htbp]
\caption{Regionalization Workflow of profile for \acs{PV} for each model.}
\label{table_PV_profiles}
\begin{tabular*}{\tblwidth}{@{}L p{15mm} p{40mm} L@{}}
 \toprule
 & \bf{\acs{BU}/ \acs{TD}} & \bf{Input source} & \bf{Allocation resolution} \\ 
 ELMOD & \acs{TD} & NUTS 0 profiles (TSOs) & Distribute into the nodes contained into the profile region \\ 
 eTraGo & \acs{BU} & Weather data \newline (ERA5 $28\times 28 \, \si{\kilo\meter}^2$) & Closest weather data location \\ 
 Europower & \acs{TD} \newline (from \acs{BU}) & NUTS 2 profiles (EMHIRES \newline using CM SAF SARAH \newline weather data, $5.55\times 5.55\, \si{\kilo\meter}^2$) & Overlap of generation regions and node regions \\ 
 ISAaR & \acs{BU} & Weather data (DWD \& \newline ECMWF-CAMS,  \newline $13\times 13\, \si{\kilo\meter}^2$;  ICON-EU) & Overlap of weather grid and NUTS 3 \\ 
 MarS\slash ZKNOT & \acs{BU} & Weather data (MERRA-2, \newline $50\times 50\, \si{\kilo\meter}^2$) & Overlap of weather grid and node regions \\ 
 MILES & \acs{BU} & Weather data \newline (COSMO-REA6, $6\times 6\, \si{\kilo\meter}^2$)  & Overlap of weather grid and node regions  \\ 
 PERSEUS & \acs{BU} & Weather data (ANEMOS, \newline$20\times 20\, \si{\kilo\meter}^2$) & Overlap of weather grid and node regions \\ 
 PowerFlex & \acs{TD} & NUTS 0 profiles \newline (TSOs) & Distribute into the nodes contained into the profile region  \\ \bottomrule
\end{tabular*}
\end{table}

Finally, Table \ref{table_PV_correction}, contains the post-processing corrections used by each model. By default, the models using a \acs{TD} approach do not need to perform corrections. The rest of the models, which use a \acs{BU} approach, harmonize the full load hours using \acs{ENTSO-E} data (\cite{entsoe_factsheets2016}). ISAaR and MILES perform correction for each month, whereas eTraGo, PERSEUS and Mars\slash ZKNOT do the corrections for the whole year.

\begin{table}[htbp, width=.6\textwidth]
\caption{Post-processing correction workflows for \acs{PV}.}
\label{table_PV_correction}
\begin{tabular*}{\tblwidth}{@{}LL@{}}
 \toprule
 & \bf{Corrections after node assignment}  \\ \midrule
 ELMOD & None \\ 
 eTraGo & NUTS 0 annual full load hours (\acs{ENTSO-E}) \\ 
 Europower & None \\ 
 ISAaR & NUTS 0 monthly full load hours (\acs{ENTSO-E}) \\ 
 MarS\slash ZKNOT &  NUTS 0 annual full load hours (\acs{ENTSO-E})
 \\ 
 MILES & NUTS 0 monthly full load hours (EEX) \\ 
 PERSEUS & NUTS 0 annual full load hours (\acs{ENTSO-E}) (optional)
 \\ 
 PowerFlex &  None \\ \bottomrule
\end{tabular*}
\end{table}

\subsubsection{Wind}\label{subsec_wind_results}

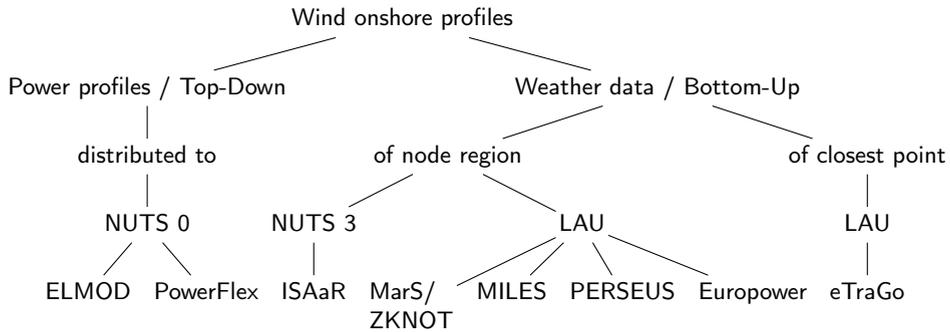
\begin{figure}[htbp]
\centering
{
\begin{forest}
for tree={align=left, inner sep=1pt, text=black, edge = {black}}
[Wind onshore profiles
    [Power profiles / Top-Down 
        [distributed to
            [NUTS 0
                [{ELMOD} ]
                [{PowerFlex} ]]]]
    [Weather data / Bottom-Up 
        [of node region
            [NUTS 3 
                [{ISAaR} ]]
            [LAU
                [{MarS\slash\\ ZKNOT} ]
                [{MILES} ]
                [{PERSEUS} ]
                [{Europower}]]]
        [of closest point
            [LAU
                [{eTraGo} ]]]]]
\end{forest}}
\caption{Dendrogram of the regionalization of wind onshore profiles workflows}
\label{forest_windon_profiles}
\end{figure}

Most models use the same workflow for the regionalization of \acs{PV} and wind onshore. However, Europower has a significant difference in its workflow: for the regionalization of wind profiles it uses a \acs{BU} approach with \acs{LAU} weather data from GWA and MERRA (with a $50\times 50\, \si{\kilo\meter}$ resolution) overlapping with the generation regions. The update of the dendrogram of workflows for wind onshore profiles can be seen in Figure \ref{forest_windon_profiles}. 

All models perform the regionalization of wind offshore similarly to the regionalization of wind onshore. However, the node allocation of the installed capacities is always performed by following the physical onshore grid connection. Hence splitting up the capacity in the North Sea and Baltic Sea, see \cite{nep2035}. Europower performs the regionalization in the same way, but adds an offshore node which gets an EMHIRES (\cite{EMHIRES}) power profile from weather data (like for \acs{PV}). Since this is still part of the grid feed-in into the optimization process, it cannot be clearly allocated to an unequivocal onshore node.

\subsubsection{Demand}

All considered models use a similar workflow for the demand regionalization, expect for PERSEUS, which builds up the profiles from \acs{LAU} level data (\acs{BU}). The remaining models follow a \acs{TD} approach, using standard demand profiles at national or federal level. The difference is mainly in the spatial resolution and the sectors considered to calculate the final profile, see Table \ref{table_demand_profiles}. The demand profiles are then altered at the node level by \textit{a demand factor} using local data of the node region in a \acs{BU} approach, see Table \ref{table_demand_cap_factor}. ELMOD and PERSEUS use a \acs{TD} approach for the demand factor at NUTS 3 level, based on population and GDP, distributed to all its nodes. PowerFlex uses a mixed approach: apart from scaling the profiles at \acs{LAU} level according to the population and energy intensive industry, the total annual demand at NUTS 1 is scaled according to statistical data taking into account the differences of other load sectors at the level of federal stats.

The post-processing correction of demand data is the same as for \acs{PV} data, Table {\ref{table_PV_correction}}, except for eTraGo, which does not perform correction at national level after the \acs{BU} approach and PERSEUS, which scale monthly demand on NUTS 0 to EUROSTAT.

\begin{table}[htbp]
\caption{Regionalization workflow for the demand profiles for each model.}
\label{table_demand_profiles}
\begin{tabular*}{\tblwidth}{@{}LLLL@{}}
 \toprule
 & \bf{\acs{BU}/ \newline \acs{TD}} & \bf{Input source} & \bf{Sectors} \\
 \midrule
 ELMOD & \acs{TD} & NUTS 0 & Total demand \\
 eTraGo & \acs{TD} & NUTS 0 & Agriculture, residential, commerce and industry \\ 
 Europower & \acs{TD} & NUTS 0 & Base load  \\ 
 ISAaR & \acs{TD} & NUTS 0 & Residential, commerce and industry \\ 
 MarS\slash \newline ZKNOT & \acs{TD} & NUTS 0 & Residential, commerce, agricultural and industry \\
 MILES & \acs{TD} & NUTS 0 & Residential and industry \\
 PERSEUS & \acs{BU} & \acs{LAU} & Residential, industry, commercial, trade and services \\ 
 PowerFlex & \acs{TD} & NUTS 0 & Total demand, but independent energy intensive industry uniform profile  \\
 \bottomrule
\end{tabular*}
\end{table}

\begin{table}[htbp]
\caption{Regionalization workflow for the demand factor for each model.}
\label{table_demand_cap_factor}
\begin{tabular*}{\tblwidth}{@{}LLLL@{}}
 \toprule
 & \bf{\acs{BU}/ \acs{TD}} & \bf{Input resolution} & \bf{Source} \\ \midrule
 ELMOD & \acs{TD} & NUTS 3 & Population, GDP \\
 eTraGo & \acs{BU} & \acs{LAU} & Population, GDP, land use \\
 Europower & \acs{BU} & \acs{LAU} & Population, temperature  \\
 ISAaR & \acs{BU} & \acs{LAU} & Population, employment \\
 MarS\slash \newline ZKNOT & \acs{BU} & \acs{LAU} & Population, GDP, temperature \\
 MILES & \acs{BU} & \acs{LAU} & Population, GDP, heat demand, conventional vehicles \\
 PERSEUS & \acs{TD} & NUTS 3 & Population, GDP, land use and temperature \\ 
 PowerFlex & \acs{BU} \& \acs{TD} & \acs{LAU} (\& NUTS 1) & Population and energy intensive industry (and NUTS 1 statistical data) \\ \bottomrule
\end{tabular*}
\end{table}

\subsection{Comparison of the regionalization output}
\label{sec_results_output}

In this section, we compare the output of the regionalization workflows for all considered models. Recall from Section \ref{sec_indicators}, that we use wind and \acs{PV} profiles, installed capacities and annual demand aggregated to NUTS 3 regions. The same computations were made for NUTS 2, NUTS 1 and NUTS 0 in order to check if the differences between models will smooth out when the region considered gets bigger. The NUTS 2 computations are in Appendix \ref{sec_NUTS2plots}, but the NUTS 1 and NUTS 0 ones are no included to avoid too much repetition.

Before we start with the NUTS 3 data, we gathered the German installed capacity and the annual demand in Table \ref{table_NUTS0_capacities} and the availability factors in Table \ref{table_NUTS0_factors}. Overall, the data for all models are very close to the \acs{ENTSO-E} Factsheets for 2016 \cite{entsoe_factsheets2016} and the \acs{ENTSO-E} Transparency website \cite{Entsoe2016_transparency}. This is to be expected, since most models either perform a post-processing or a \acs{TD} approach (Table \ref{table_PV_correction}). Notice that, Europower (which uses data for 2015) has significantly more PV and wind onshore installed capacity than the rest of the models, but the availability factors are close to the other models. ISAaR has less wind onshore and wind offshore installed capacities, but then it has higher \acs{PV} and wind offshore availability factors, because it uses input data from 2012 scaled to get the full load hours for 2016. There are two blocks regarding annual load: ISAaR, MarS\slash ZKNOT, MILES and PowerFlex with an approximate annual load of \SI{550}{\tera\watthour} and the rest of the models with an approximate annual load of \SI{510}{\tera\watthour} or less.

Maps of installed capacities and annual demand of all models in NUTS 3 level were generated and are plotted in Appendix \ref{sec_NUTS3plots}.

\begin{table}[htbp, width=.85\textwidth]
\caption{Installed capacities and annual demand for Germany in 2016.}
\label{table_NUTS0_capacities}
\begin{tabular*}{\tblwidth}{@{}LLLLL@{}}
 \toprule
 & \bf{PV} (\si{\giga\watt}) & \bf{Wind onshore} (\si{\giga\watt}) & \bf{Wind offshore} (\si{\giga\watt}) & \bf{Annual load} (\si{\tera\watthour}) \\ \midrule
 ELMOD & 39.79 &	45.00 &	4.12 &	511.53
 \\ 
 eTraGo & 38.51 &	41.29 &	3.36 &	505.15
 \\ 
 Europower & 44.12 & 57.67 &	3.90 &	500.06
  \\ 
 ISAaR & 39.11	& 35.00 &	2.48 &	554.71
 \\ 
 MarS\slash ZKNOT & 39.79 &	45.00 &	4.12 &	556.10
 \\ 
 MILES & 39.79 &	45.00 &	4.15 &	554.66
 \\ 
 PERSEUS & 40.72 &	45.45 &	4.13 &	511.24
 \\ 
 PowerFlex & 39.89 &	45.12 &	4.13 &	557.17
 \\ 
 \acs{ENTSO-E} Factsheets (\cite{entsoe_factsheets2016}) &  39.79 &	45.00 &	4.12 &	548.40
 \\ 
 \acs{ENTSO-E} Transparency (\cite{Entsoe2016_transparency}) &  38.69 &	41.17 &	3.28 &	481.09
 \\ \bottomrule
\end{tabular*}
\end{table}

\begin{table}[htbp, width=.7\textwidth]
\caption{Availability factors for Germany in 2016.}
\label{table_NUTS0_factors}
\begin{tabular*}{\tblwidth}{@{}LLLL@{}}
 \toprule
 & \bf{PV} & \bf{Wind onshore} & \bf{Wind offshore} \\ \midrule
 ELMOD & 0.101 &	0.169 &	0.332
 \\ 
 eTraGo & 0.102 &	0.179 &	0.322
 \\ 
 Europower & 0.100 &	0.167 &	0.334
  \\ 
 ISAaR & 0.114 &	0.163 &	0.570
 \\ 
 MarS\slash ZKNOT & 0.100 &	0.157 &	0.334
 \\ 
 MILES & 0.100	& 0.176 &	0.370
 \\ 
 PERSEUS & 0.100 &	0.179 &	0.340
 \\ 
 PowerFlex & 0.100 &	0.167 &	0.334
  \\ 
 \acs{ENTSO-E} Factsheets (\cite{entsoe_factsheets2016}) &  0.100 &	0.167 &	0.334
 \\ 
 \acs{ENTSO-E} (\acs{OPSD} \cite{OPSD-project-website}) & 0.100 &	0.184 &	0.328
 \\ \bottomrule
\end{tabular*}
\end{table}

\subsubsection{Photovoltaics}\label{subsec_PV_results_output}

\begin{figure}[htbp]
\centering
\includegraphics[width=0.75\linewidth]{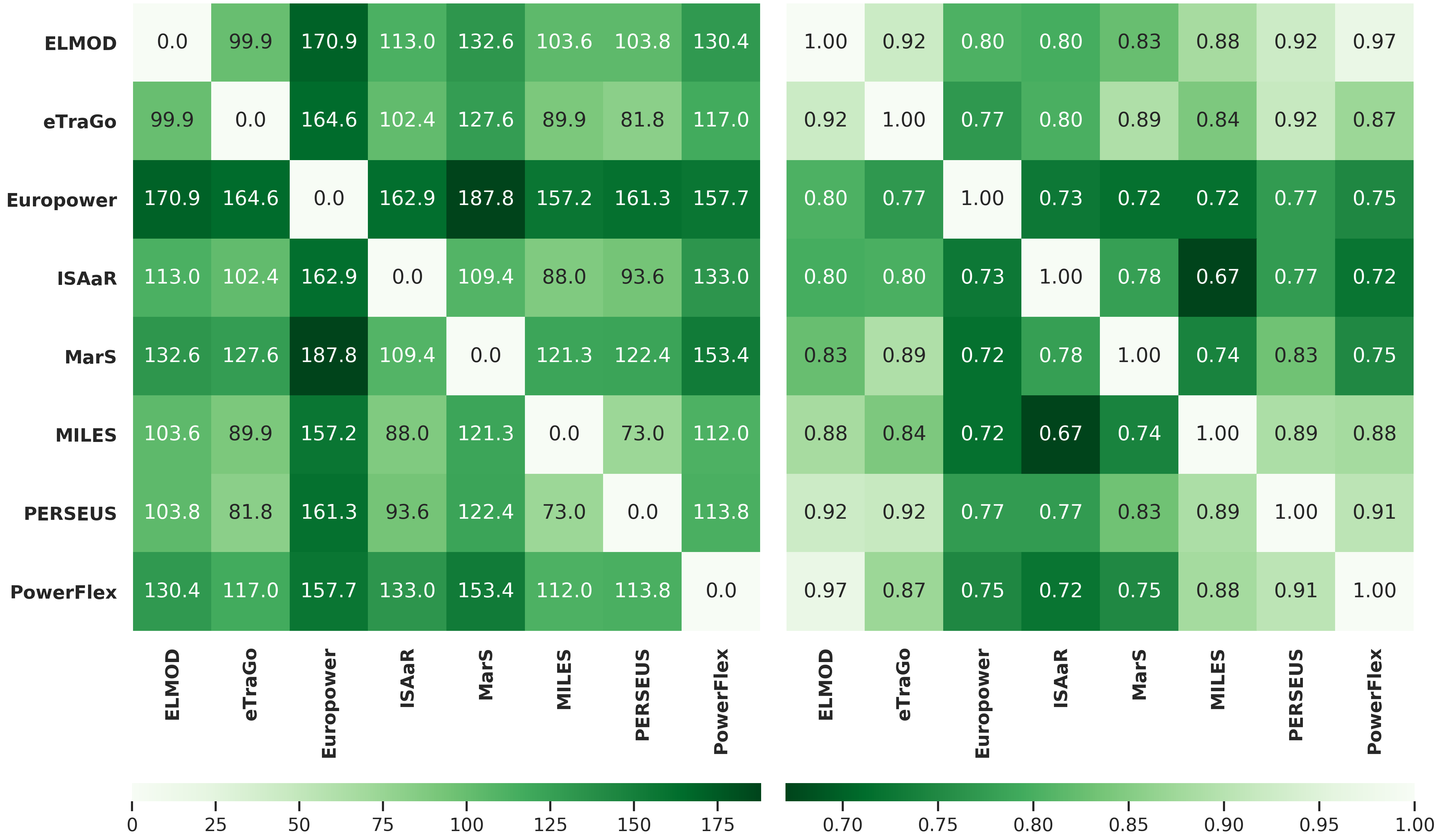}
\caption{\acs{RMSE} of \acs{PV} installed capacities differences ({\si{\mega\watt}}) across all NUTS 3 regions (left) and mean Pearson correlation of \acs{PV} profiles across intersecting NUTS 3 regions (right).}
\label{fig_PV_inst_cap_Pearson_matrices}
\end{figure}

The comparison of the regionalization output for \acs{PV} is calculated using the \acs{RMSE} of installed capacities differences across all NUTS 3 regions and the mean of the Pearson correlation of the profiles for every pair of models across all regions with installed capacity (see Section \ref{sec_indicators} for details). The results are shown in the two symmetric matrices in Figure \ref{fig_PV_inst_cap_Pearson_matrices}.

For the installed capacity, Europower has the most different installed capacity regionalization. In Figure \ref{fig_PV_NUTS3}, it is clear that the installed capacity is differently distributed in contrast to all the other models, with a higher concentration in the Brandenburg region. At national level, Europower calculates a much higher installed capacity (Table \ref{table_NUTS0_capacities}). After Europower, PowerFlex and MarS\slash ZKNOT have also a more different regionalization outcome for PV. For Europower and PowerFlex their singularity can be explained as both use a {\acs{TD}} approach, see Section {\ref{forest_PV_installed}}. Concerning the rest of the models, MILES and PERSEUS are the models with a closer \acs{PV} installed capacity regionalization.

In general, all the models have a similar \acs{PV}  regionalization profile, although Europower and ISAaR are using a different profile year. That is explained by the daily pattern of the sun: all \acs{PV} profiles should be an alteration of the same clear sky profiles, which would only differ on the geographical location. By aggregating the data per NUTS 2 (see Figure \ref{fig_PV_inst_cap_Pearson_matrices_nuts2}), NUTS 1 and NUTS 0 all model profiles are very close to each other with Europower, ISAaR and MarS\slash ZKNOT showing little differences. It is also remarkable that, the pair of models with closer profile regionalization are the two using a top-down approach with NUTS 0 profiles: ELMOD and PowerFlex, see Figure {\ref{forest_PV_profiles}}.

\subsubsection{Wind onshore}
\label{subsec_windonshore_results_output}

For wind onshore installed capacities, Europower uses a different regionalization than the rest of the models, having the highest total installed capacity, see Table \ref{table_NUTS0_capacities}. This might be due to the different weather year used and the indirect calculation of the installed capacities using the full load hours (see Section \ref{sec_indicators}).

The plots in Figure \ref{fig_windonshore_NUTS3} show most of the wind onshore installed in the north for all models, since there is higher wind availability. However, there are differences regarding the distribution in the northern regions, especially for the states of Niedersachsen, Schleswig-Holstein, Mecklenburg-Vorpommern and Brandenburg. In the cases under study, wind onshore regionalization produces more differentiated output than \acs{PV}. We can identify a group with a slightly closer wind onshore installed capacity regionalization formed by eTraGo, ISAaR, PERSEUS and PowerFlex.

\begin{figure}[htbp]
\centering
\includegraphics[width=0.75\linewidth]{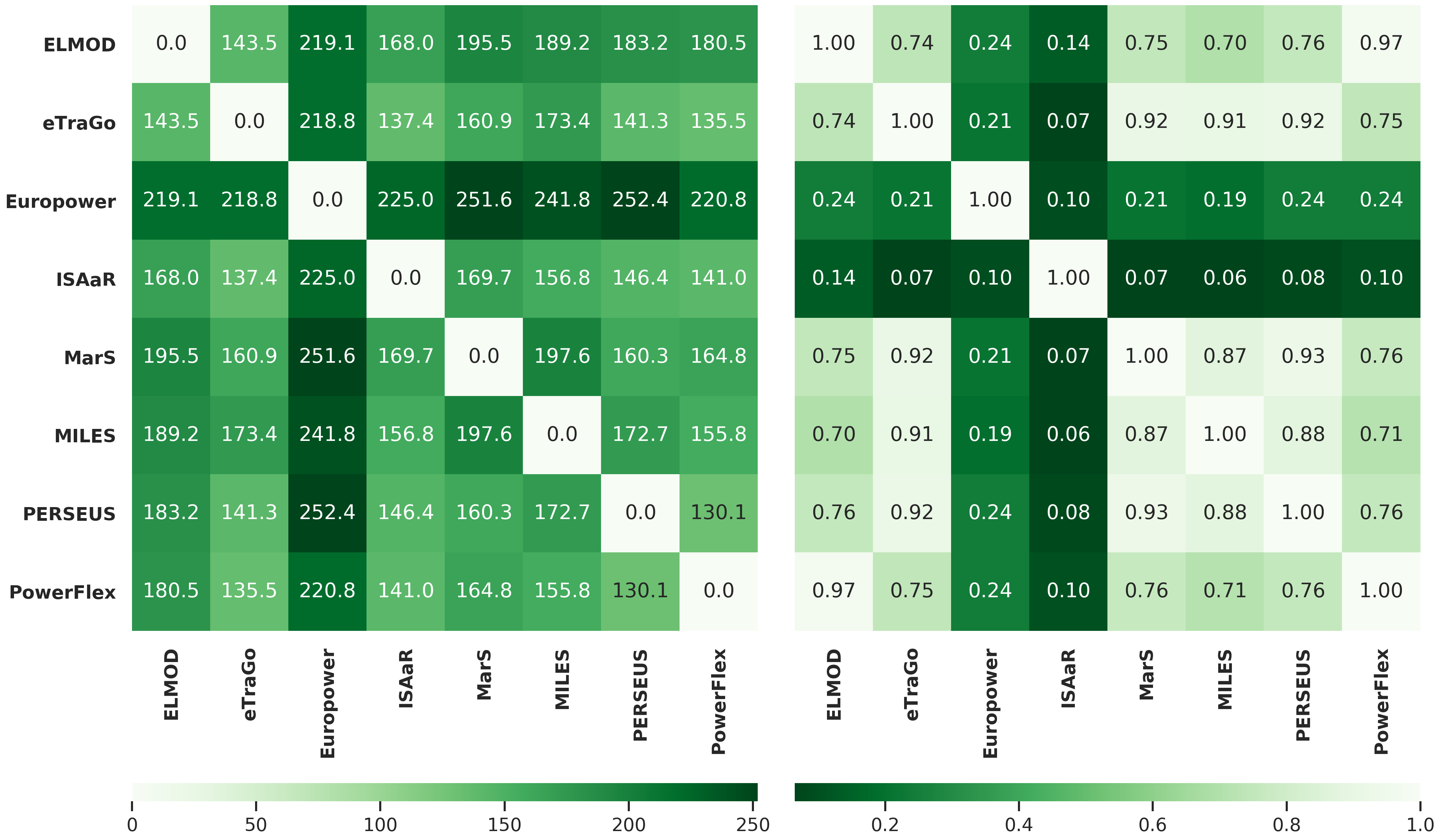}
\caption{\acs{RMSE} of wind onshore installed capacities differences ({\si{\mega\watt}}) across all NUTS 3 regions (left) and mean Pearson correlation of wind onshore profiles across intersecting NUTS 3 regions (right).}
\label{fig_windonshore_inst_cap_Pearson_matrices}
\end{figure}

Europower and ISAaR have a very different regionalization profiles than the rest of the models (which was only slightly seen in the \acs{PV} profile comparison, Figure \ref{fig_windonshore_inst_cap_Pearson_matrices}). This difference persists and is increased if one aggregates at NUTS 2, where all the reginoalization of the profiles are very close except for these two models, see Figure \ref{fig_windonshore_inst_cap_Pearson_matrices_nuts2}. This was expected, since both models use different weather years than the other models. Notice also that, the rest of the models have a close profile regionalization and two groups can be identified: PowerFlex and ELMOD on one side and eTraGo, MarS\slash ZKNOT, MILES and PERSEUS on the other side. The fact that PowerFlex and ELMOD have a similar profile regionalization output can be explained by the fact that they both use a in a \acs{TD} approach with NUTS 0 profiles.

\subsubsection{Wind offshore}
\label{subsec_windoffshore_results_output}

\begin{figure}[htbp]
\centering
\includegraphics[width=0.75\linewidth]{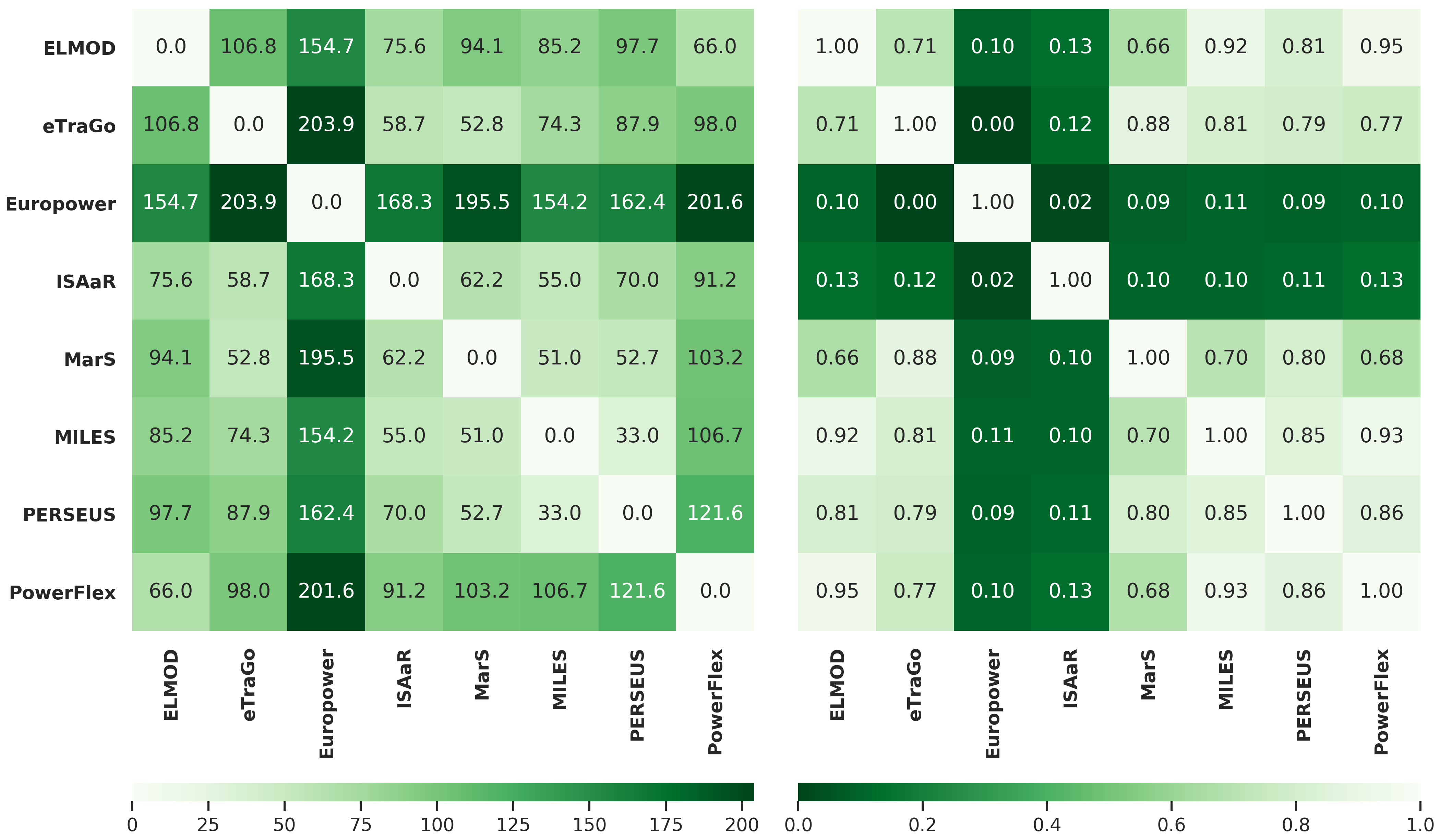}
\caption{\acs{RMSE} of wind offshore installed capacities differences ({\si{\mega\watt}}) across all NUTS 3 regions (left) and mean Pearson correlation of wind offshore profiles across intersecting NUTS 3 regions (right).}
\label{fig_windoffshore_inst_cap_Pearson_matrices}
\end{figure}

As in the case of \acs{PV} and wind onshore, Europower has a different wind offshore installed capacity regionalization output (see Figure \ref{fig_windoffshore_inst_cap_Pearson_matrices}), since it used 2015 data, also because the offshore grid nodes were assigned to the closest onshore NUTS 3 region for the comparison in this paper. As for the rest of the models, the two models using a \acs{TD} approach (ELMOD and PowerFlex) have the closest output and are quite different from the rest of the models output. Nevertheless, Figure \ref{fig_windoffshore_NUTS3} shows that the onshore allocation of wind offshore data us very similar, since all models use the real grid connections to the shore.

\begin{figure}[htbp]
\centering
\includegraphics[width=0.75\linewidth]{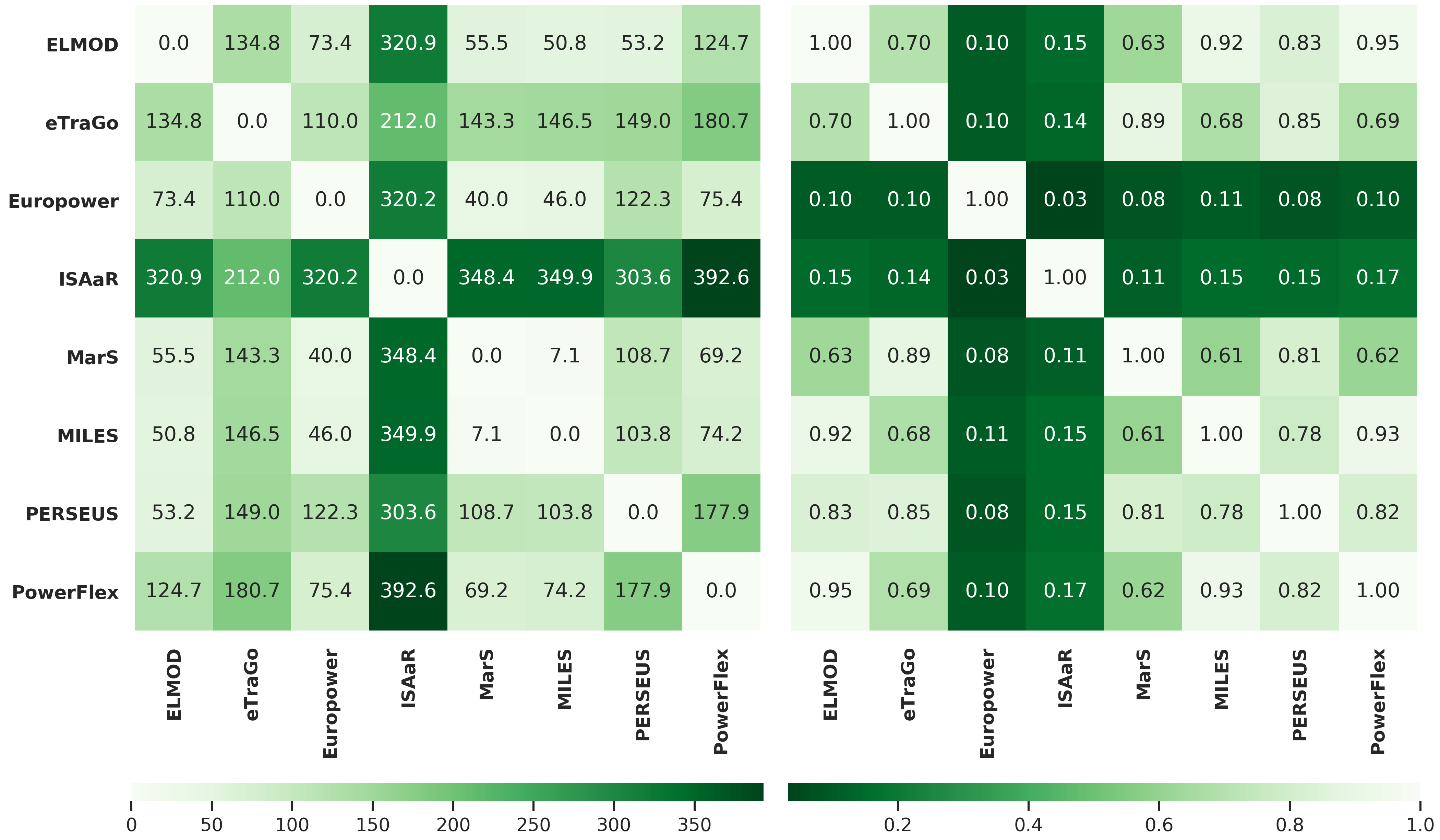}
\caption{\acs{RMSE} of wind offshore installed capacities differences ({\si{\mega\watt}}) of the Noth sea and the Baltic sea (left) and mean Pearson correlation of the North sea and the Baltic sea wind offshore profiles (right).}
\label{fig_windoffshore_inst_cap_Pearson_matrices_nordostsee}
\end{figure}

The matrices comparing wind offshore regionalization using only two regions, the North Sea and the Baltic Sea can be found in Figure \ref{fig_windoffshore_inst_cap_Pearson_matrices_nordostsee}. We see important differences in the installed capacity comparison. This is because our method is designed for comparing regionalization among many regions. When comparing two regions, the outcome is mainly determined by the differences at national level, where especially ISAaR and eTraGo have a small installed capacity with respect to the others (Table \ref{table_NUTS0_capacities}). MILES and MarS\slash ZKNOT have very close values (see Figure \ref{fig_windoffshore_inst_cap_Pearson_matrices_nordostsee}) because they have the same installed capacity at national level and a very similar distribution between the two areas consider (see Table \ref{table_nordostsees_capacities}). Therefore, for comparing the installed capacity in only two regions, the methods used in this paper are excessively sophisticated and the information in Table \ref{table_nordostsees_capacities} is enough. However, this simplification does not allow a full comparison of the wind offshore regionalization which considers the whole grid into account.

Note that, the profile comparison matrices are very close in both comparisons (Figure \ref{fig_windoffshore_inst_cap_Pearson_matrices} and Figure \ref{fig_windoffshore_inst_cap_Pearson_matrices_nordostsee}). Europower and ISAaR have significantly more distant profile regionalization outputs, which is explained by the fact that they use different weather years. The rest of the models have a similar regionalization, especially PowerFlex and ELMOD which are the two models using a \acs{TD} approach.

\begin{table}[htbp, width=.8\textwidth]
\caption{Share of wind offshore installed capacities in the Baltic sea.}
\label{table_nordostsees_capacities}
\begin{tabular*}{\tblwidth}{ @{}LLLLLLLL@{}}
 \toprule 
 ELMOD & eTraGo & Europower & ISAaR & MarS\slash ZKNOT & MILES & PERSEUS & PowerFlex \\ \midrule
 0.10 &	0.20 &	0.13 &	0.08	& 0.09 &	0.13 &	0.10 &	0.17 \\ 
 \bottomrule
\end{tabular*}
\end{table}

\subsubsection{Demand}
\label{subsec_load_results_output}

\begin{figure}[htbp]
\centering
\includegraphics[width=0.75\linewidth]{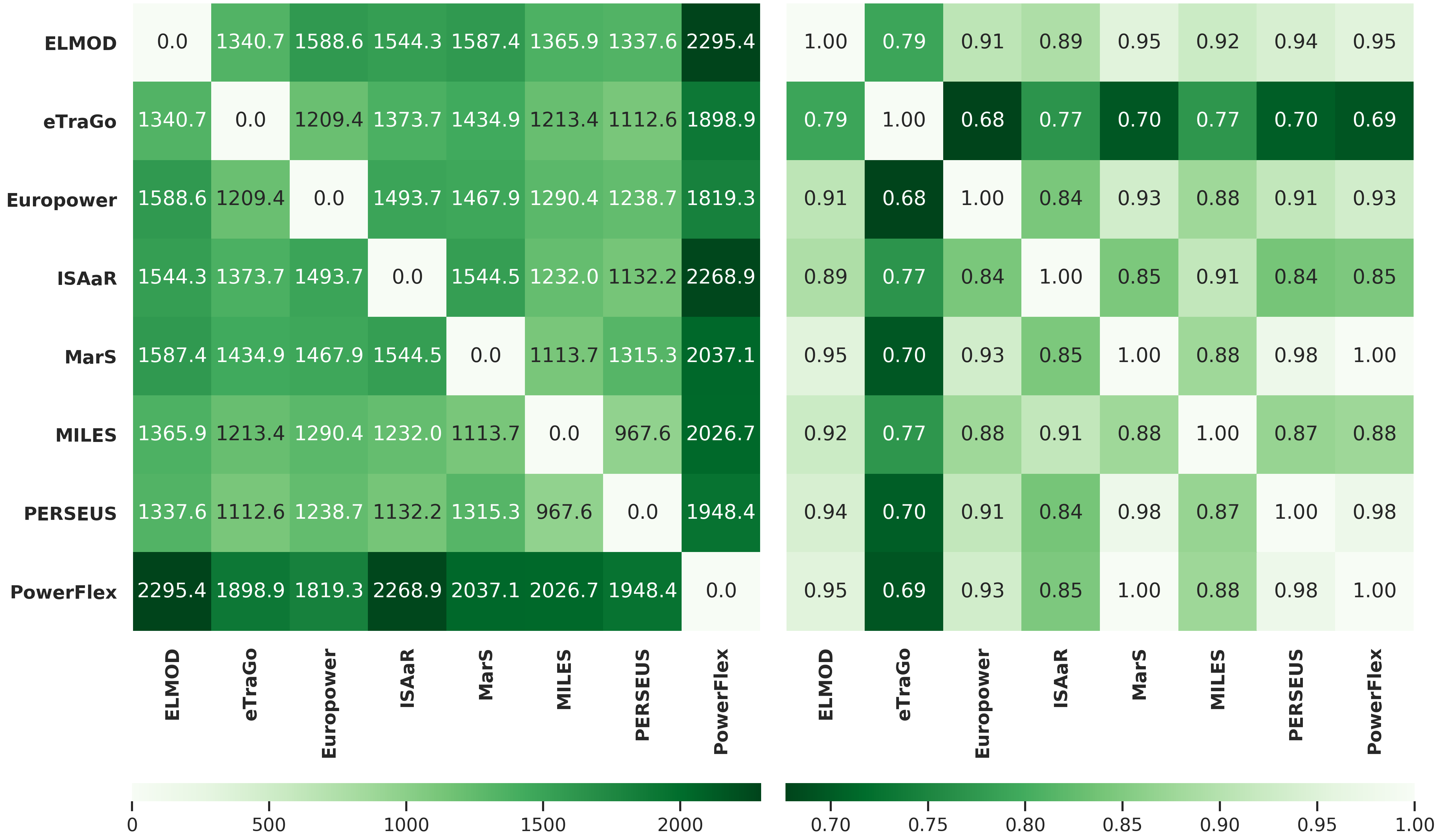}
\caption{\acs{RMSE} of annual load ({\si{\giga\watthour}}) differences across all NUTS 3 regions (left) and mean Pearson correlation of load profiles across intersecting NUTS 3 regions (right).}
\label{fig_load_inst_cap_Pearson_matrices}
\end{figure}

Regarding annual load, Figure \ref{fig_load_inst_cap_Pearson_matrices} shows that PowerFlex has the most different annual load regionalization output. This can be explained by the different workflow used for the regionalization as Figure \ref{table_demand_cap_factor} illustrates. The total demand is taken from the literature not only at the national level, but also at the federal level and constant load profiles at \acs{LAU} level for intensive power plants are included. MILES and PERSEUS are the models with a closest regionalization of demand factor, although not significantly different than the rest of the models. On the whole, annual demand regionalization output seems to be different even if models use mainly the same workflow. This suggests that, in this case, the chosen data source and the demand sectors considered play a more important role.

The profiles regionalization output show a big of similarity, especially between PERSUS, PowerFlex and MarS\slash ZKNOT models. This may be explained by the weekly pattern of the load and that most models use NUTS 0 standard load profiles. This can be also noticed by comparing Figure \ref{fig_load_inst_cap_Pearson_matrices} (NUTS 3) and Figure \ref{fig_load_inst_cap_Pearson_matrices_nuts2} (NUTS 2), where profile comparison remains almost the same, while the annual load comparison has significant differences. Notice also that, the models adapted from a different load year (eTraGo, Europower, ISAaR and MILES) show stronger differences. Nevertheless, the profile regionalization comparison shows that, adapting the profiles from different years to start in the same day of the week helps making them more comparable.

\section{Discussion}
\label{sec_discussion}

\begin{figure}[htbp]
\centering
\includegraphics[width=0.55\linewidth]{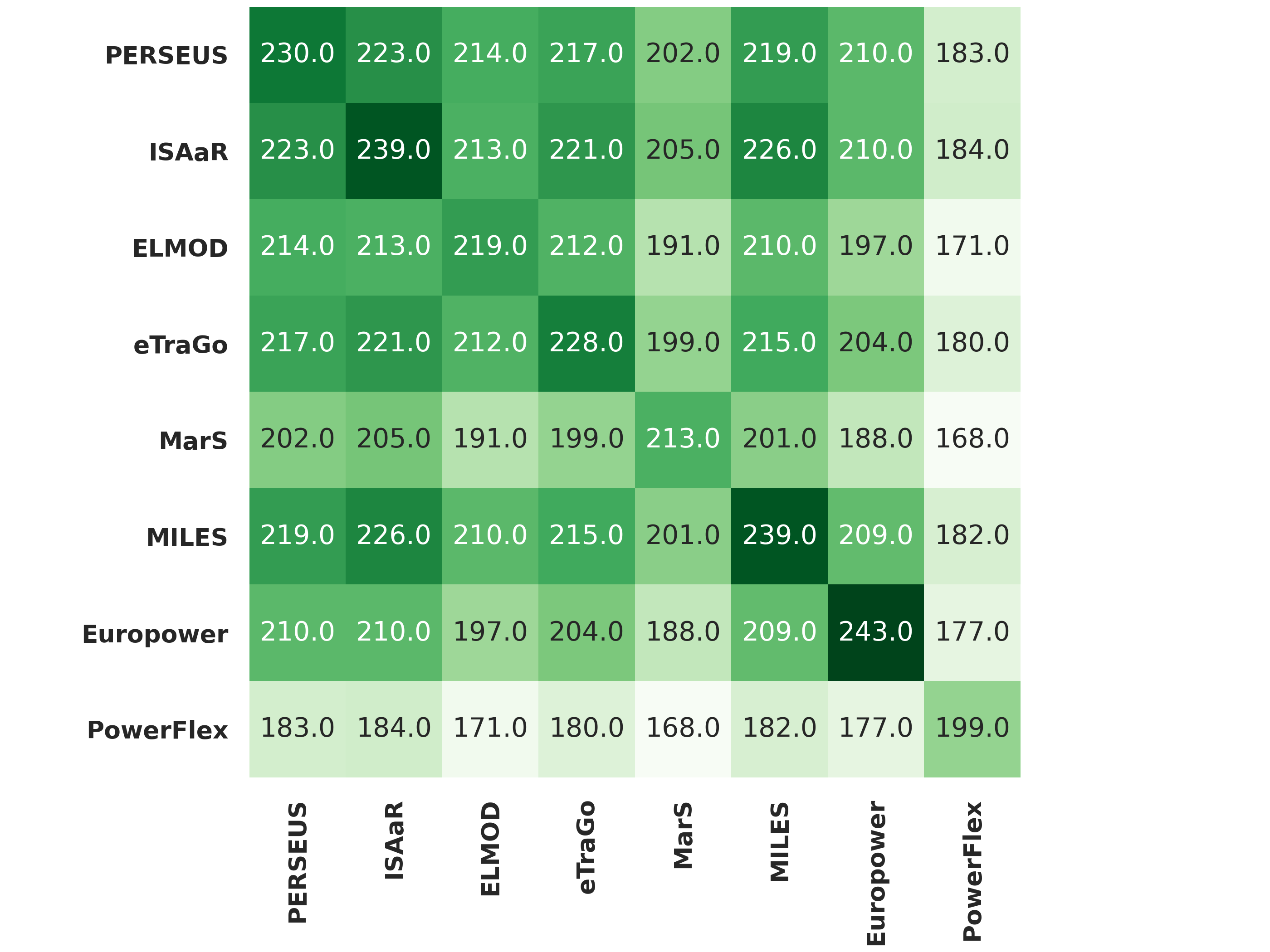}
\caption{Number of intersections between models.}
\label{fig_intersection_load}
\end{figure}

Although we only included the comparison at NUTS 3 level in Section \ref{sec_results_output} and in NUTS 2 in Appendix \ref{sec_NUTS2plots}, we also calculated it at NUTS 1 and NUTS 0 levels. Naturally, by reducing the spatial resolution of the regionalization of the installed capacity, the outputs become closer, especially for \acs{PV}. The same happens with profiles, except for the models having input data from different years, which keep being very different. This shows that, to be able see the effects of the regionalization on the transmission grid, data from NUTS 3 level or smaller is required, otherwise the differences are smoothed out. This also became clear in Section \ref{subsec_windoffshore_results_output}, where the comparison of the regionalization of installed capacities using just two wind offshore regions made the regionalization output for all models seem closer. In addition, the big overlap of regions that contain an ehv-node with load, as can be seen in Figure \ref{fig_intersection_load}, favors the use of NUTS 3 regions for the regionalization. Over a total of \num{401} NUTS 3 regions, the models have between 199 (PowerFlex) and 243 (Europower) NUTS 3 regions with \acs{ehv-nodes}, with a high number of intersecting regions. This is not counter-intuitive since all models attempt to model the same transmission grid. Looking at the results, we see no correlation between the number of intersection of regions with \acs{ehv-nodes} and the regionalization output comparison results. For instance, the fact that MarS\slash ZKNOT or PowerFlex have less NUTS 3 regions with an ehv-node does not directly imply that their regionalization is completely different from the other models. The fact that MILES and ISAaR have the highest number of intersecting NUTS 3 regions with \acs{ehv-nodes} does not imply neither that they have the closest regionalization output. This confirms that our methodology can effectively evaluate the difference in regionalization outputs despite not having installed capacities and load in the exact the same regions.

Overall, the regionalization of \acs{PV} results in the most similar output independent from the regionalization workflow. This can be explained by the strong daily pattern of \acs{PV} generation and using the installed capacity at a national level (often fixed either by the post-processing or by the use of a \acs{TD} approach). This brings the regionalization of the different models remarkably close to each other, even if weather data from different years is used (see Figure \ref{fig_PV_inst_cap_Pearson_matrices}).

The difference between \acs{TD} and \acs{BU} workflows of \acs{RES} is captured by the invariants proposed in this article, which can be seen by comparing the results in Section {\ref{sec_reg_process}} and Section {\ref{sec_results_output}}. This is more clearly visible in the regionalization of wind compared to the \acs{PV}. For demand, since all models have a similar regionalization workflow, the output of the regionalization seems to be more sensitive to the data source of demand factors and profiles.

As explained in Section \ref{sec_indicators}, the installed capacities for PowerFlex, Europower and ISAaR are calculated using \acs{ENTSO-E} full load hours for 2016. This allows for the Europower and ISAaR data to be more comparable to models using the same weather year. This confirms that the weather year is a determinant factor for the comparison of \acs{RES} regionalization output. The importance of the weather year choice is much more visible in the regionalization of wind, because it does not have a regular pattern. At a national level, we observe that Europower has more wind onshore installed capacity than the other models and ISAaR has less wind onshore and wind offshore installed capacities (see Table \ref{table_NUTS0_capacities}). A possible explanation might be that, by calculating the installed capacities using profiles from 2015 (Europower) and 2012 (ISAaR), but full load hours from 2016, the installed capacities get distorted by the difference in wind full load hours of those years. Regarding the demand however, the shifting of the data such that it starts at the same day of the week, brings the profiles comparison close together.

\section{Conclusion}
\label{sec_conclusion}

In this paper, we developed a methodology to compare regionalization techniques of input data for \acs{PV}, wind and electrical demand between various models as well as the assignment of regionalized data to \acs{ehv-nodes}. In a first step, we reduced the complexity of regionalization workflows by dividing the process into the regionalization of the data and the grid node assignment. Comparing the strategies of the models under analysis, we were able to display the main differences in the workflows as simple dendrograms.
In a second step, we defined two invariants to evaluate the outcome of the regionalization process at the NUTS 3 level, one invariant for annual profiles (including demand and \acs{RES} availability) and one for installed capacities and annual demand. This methodology enabled us to illustrate the comparison of very different regionalization workflows and its outcome to simple parameters easy to visualize. Our results show that the resolution of the input data and the use of \acs{TD} or \acs{BU} approaches are the most determinant factors in the regionalization process. We found out that NUTS 3 is an adequate spatial resolution level for comparing regionalization methods for transmission grid modeling. A coarser resolution would not allow for detecting the differences between the regionalization processes and their output.

Taking into account the variety of different regionalization workflows, it is remarkable that most models end up with very similar profile regionalization output. This is especially true with \acs{PV}, because of its daily pattern, but also for the demand, because of its weekly pattern and also because most models use NUTS 0 standard load profiles as input.

The methodology presented in this paper has also some limitations. For instance, it is not adequate to compare the regionalization of installed capacities from different years, especially regarding the profile regionalization. For load profiles, however, modifying the profiles so that they all start in the same day of the week, make the load profile regionalization more comparable.

\subsection*{Acknowledgments}
The authors like to acknowledge the contribution of all the partners in the MODEX-Net project who either provided the source data or commented on earlier drafts of this paper. Concretely, from FZJ-IEK3: Jochen Linssen and Chloi Syranidou; from DLR-VE: Julian Bartels and Jan Buschmann; from FfE: Timo Kern, Felix B\"{o}ing, Tobias Schmid and Christoph Pellinger; from KIT-IIP: Armin Ardone, Katrin Seddig and Manuel Ruppert; from \"{O}ko-Institut e.V., Energy \& Climate Division: Matthias Koch, Christoph Heinemann, Franziska Flachsbarth, Sebastian Palacios, Susanne Krieger and Christian Winger; from RWTH Aachen University-IAEW: Jonas Mehlem, Lukas Weber, Levin Skiba, Alexander Fehler and Annika Klette; from TU Dortmund (ie3): Bj\"{o}rn Matthes and Jan Peper; and from TU Dresden, Lehrstuhl f\"{u}r Energiewirtschaft: Hannes Hobbie, Christina Wolff, David Sch\"{o}nheit and Dominik M\"{o}st.

\printcredits

\bibliography{references}

\newpage
\appendix
\section{Comparison of regionalized RES feed-in and demand}\label{sec_NUTS3plots}

\begin{figure}[htbp]
\centering
\includegraphics[width=\textwidth]{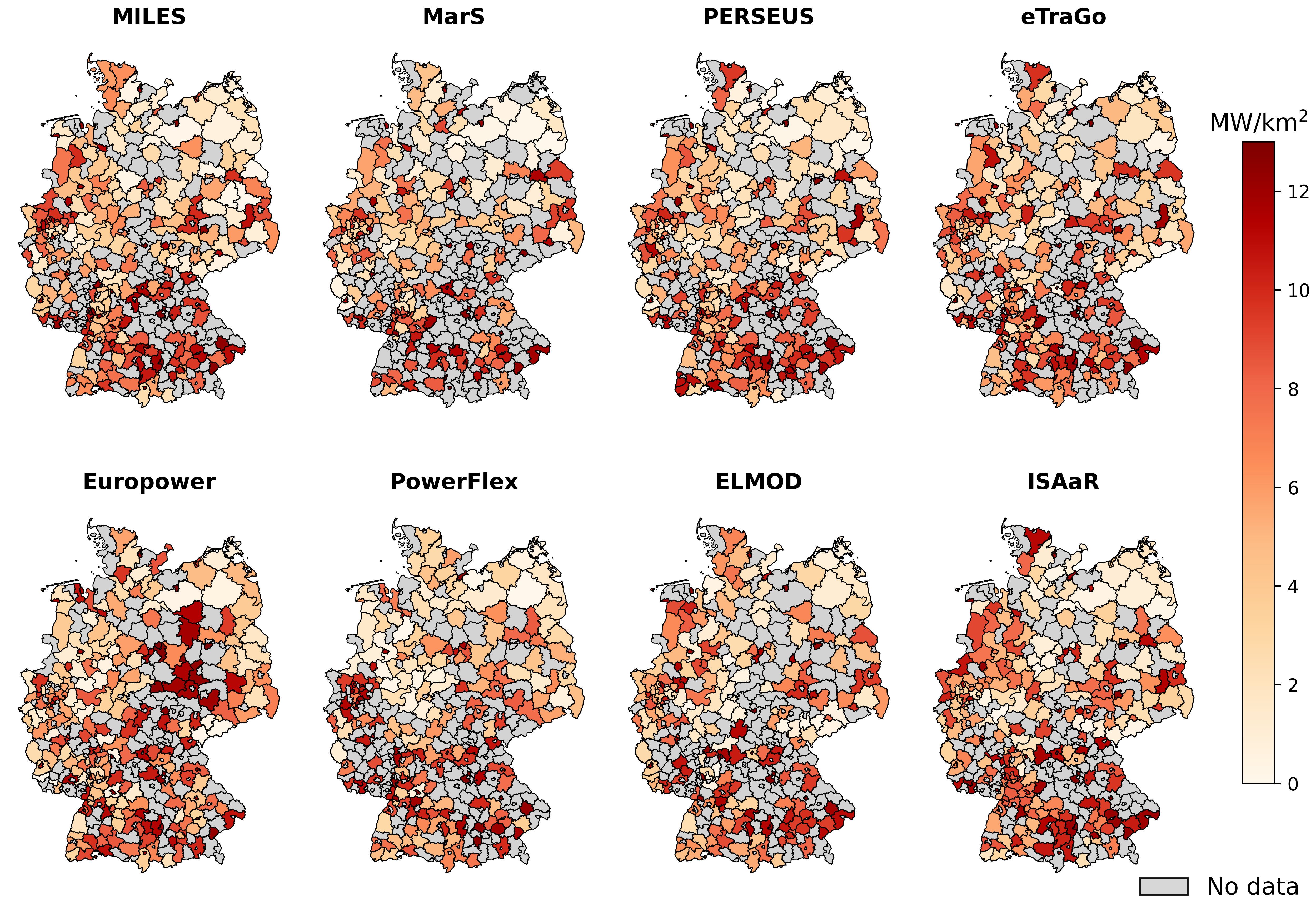}
\caption{PV installed capacity at \acs{ehv-nodes} aggregated per NUTS 3 regions density plot in {\si{\mega\watt\per\kilo\meter\squared}}.}
\label{fig_PV_NUTS3}
\end{figure}

\begin{figure}[htbp]
\centering
\includegraphics[width=\textwidth]{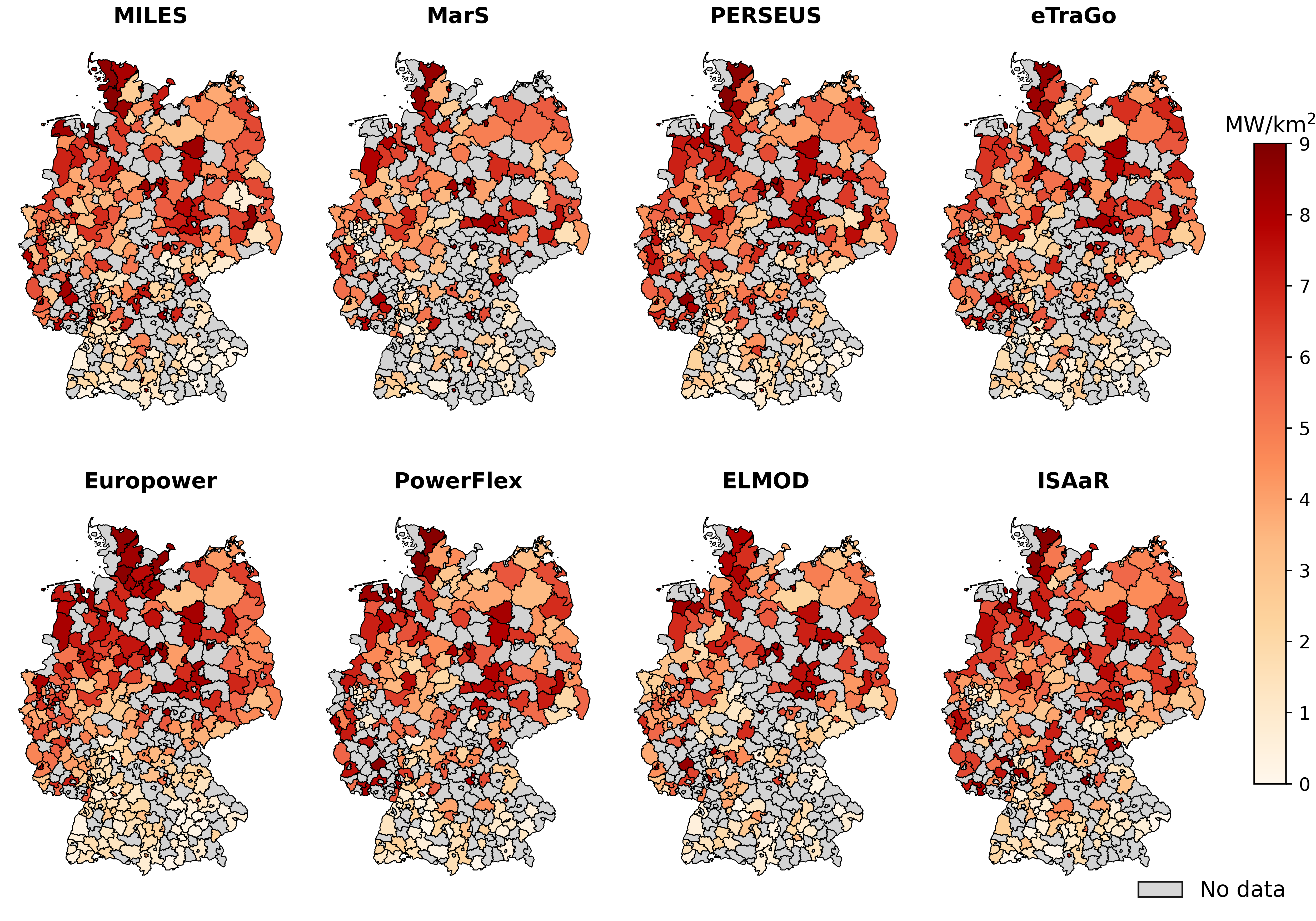}
\caption{Wind onshore installed capacity at \acs{ehv-nodes} aggregated per NUTS 3 regions density plot in {\si{\mega\watt\per\kilo\meter\squared}}.}
\label{fig_windonshore_NUTS3}
\end{figure}

\begin{figure}[htbp]
\centering
\includegraphics[width=\textwidth]{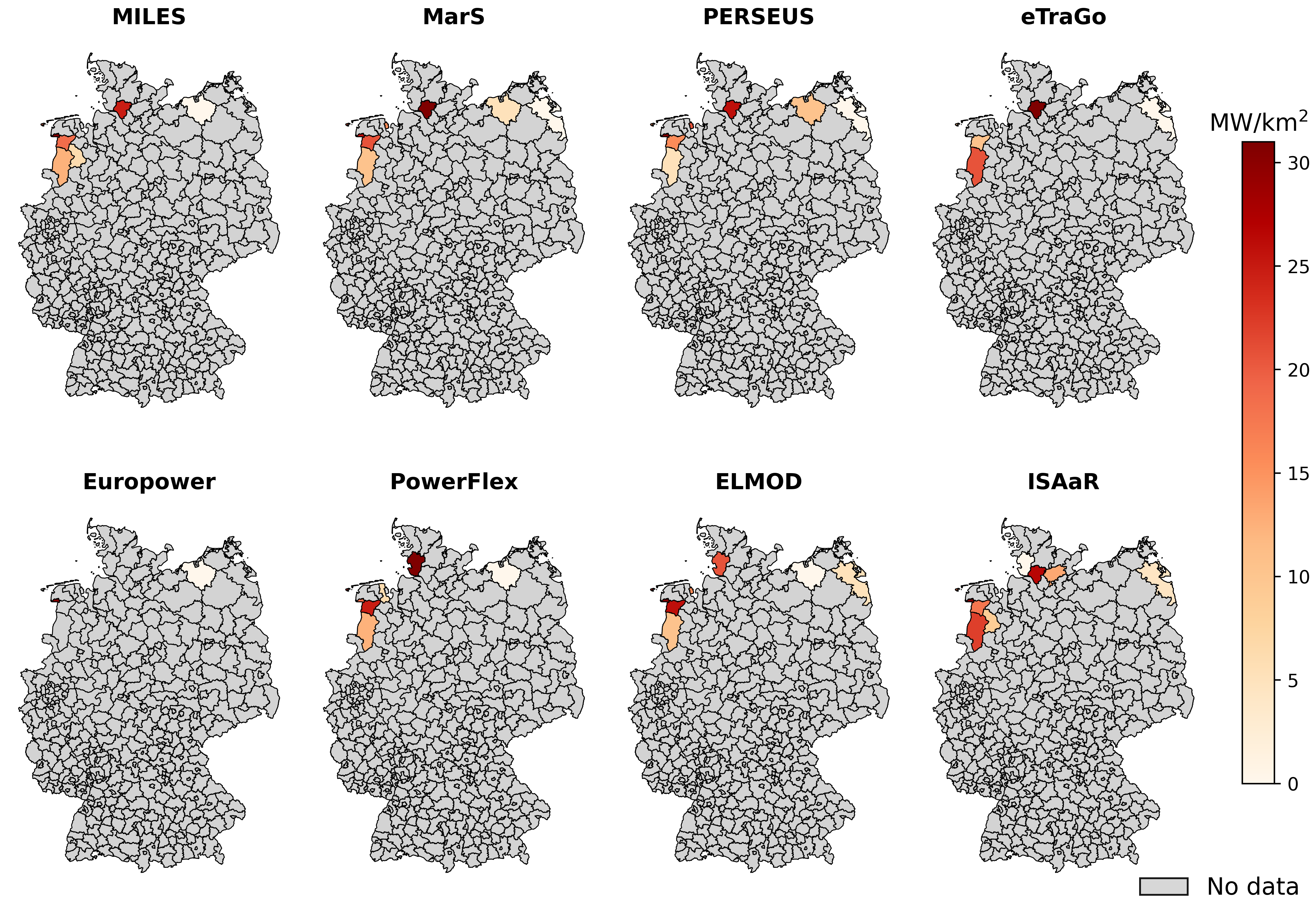}
\caption{Wind offshore installed capacity at \acs{ehv-nodes} aggregated per NUTS 3 regions density plot in {\si{\mega\watt\per\kilo\meter\squared}}.}
\label{fig_windoffshore_NUTS3}
\end{figure}

\begin{figure}[htbp]
\centering
\includegraphics[width=\textwidth]{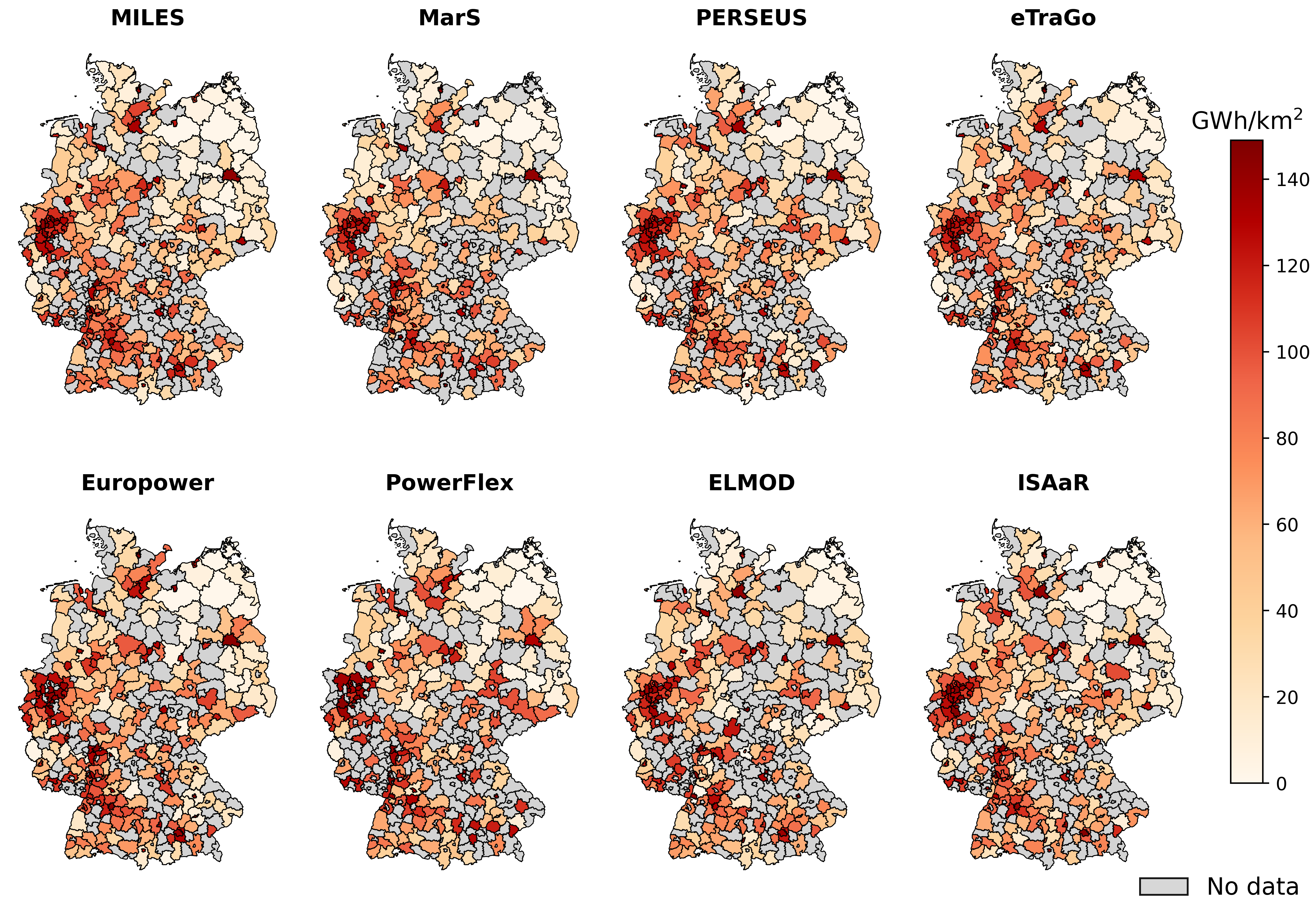}
\caption{Annual demand installed capacity at \acs{ehv-nodes} aggregated per NUTS 3 regions density plot in {\si{\giga\watthour\per\kilo\meter\squared}}.}
\label{fig_load_NUTS3}
\end{figure}

Figure \ref{fig_PV_NUTS3}, Figure \ref{fig_windonshore_NUTS3}, Figure \ref{fig_windoffshore_NUTS3} and Figure \ref{fig_load_NUTS3} depict the installed capacities of \acs{PV}, wind onshore, wind offshore in \si{\mega\watt\per\kilo\meter\squared} and annual electricity demand in \si{\giga\watthour\per\kilo\meter\squared}, respectively, for all models under analysis summing up the data of all \acs{ehv-nodes} for each NUTS~3 region. Gray colored regions do not contain any \acs{ehv-nodes}, thus no demand or \acs{RES} feed-in is assigned to that region. This does not mean that in the gray regions there is no installed capacity or demand.

\clearpage

\section{Comparison at NUTS 2 level}\label{sec_NUTS2plots}

In this section, we add the matrices showing the RMSE and the Pearson correlation comparing annual demand, installed capacities and profiles at NUTS 2 level. These matrices support the fact that the NUTS 3 level is necessary to see some finer distinctions between model regionalizations, but also show some convergence in the differences between the models when the resolution is reduced.

\begin{figure}[htbp]
\centering
\includegraphics[width=0.75\linewidth]{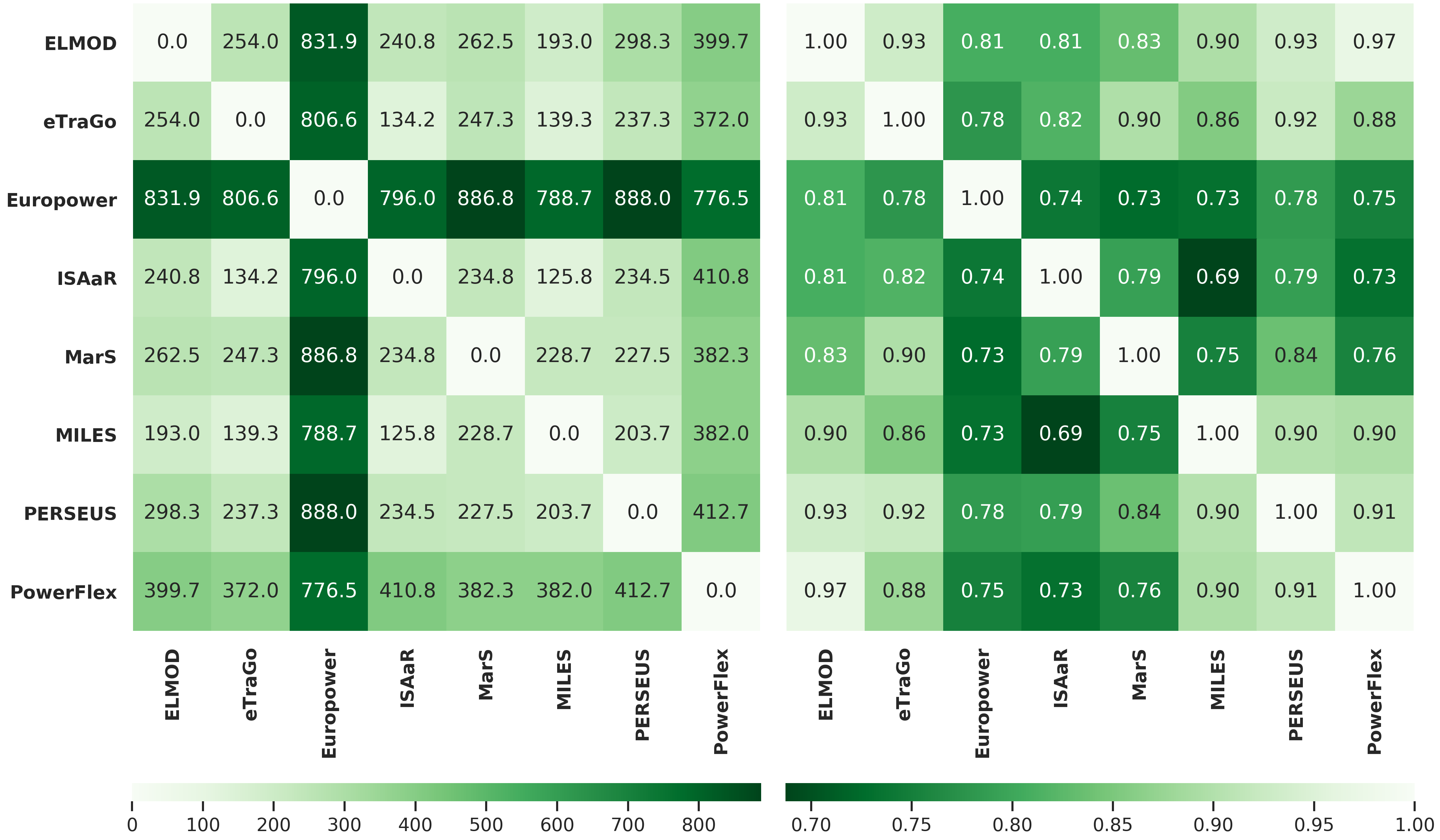}
\caption{\acs{RMSE} of \acs{PV} installed capacities differences ({\si{\mega\watt}}) across all NUTS 2 regions (left) and mean Pearson correlation of \acs{PV} profiles across intersecting NUTS 2 regions (right).}
\label{fig_PV_inst_cap_Pearson_matrices_nuts2}
\end{figure}

\begin{figure}[htbp]
\centering
\includegraphics[width=0.75\linewidth]{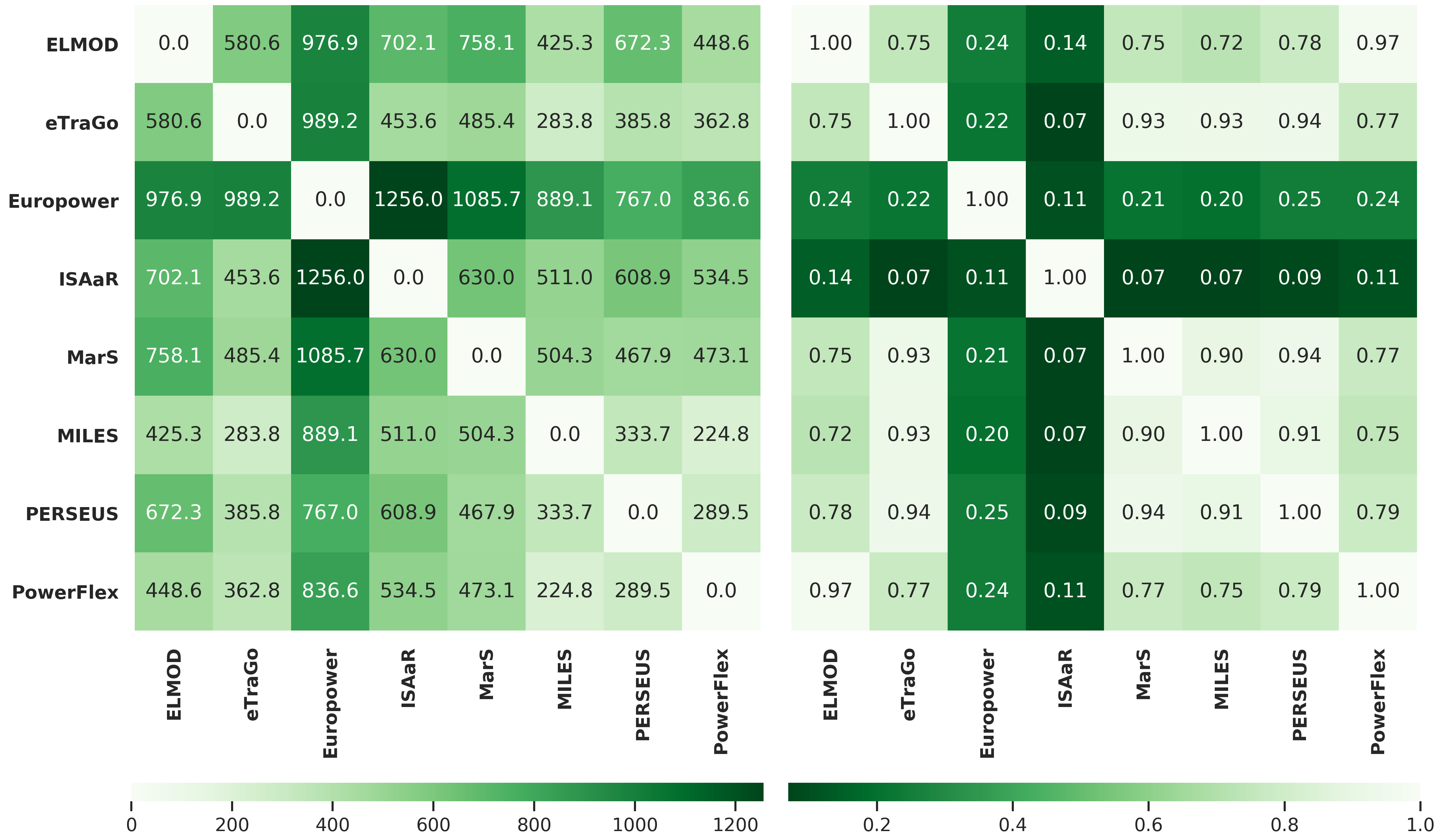}
\caption{\acs{RMSE} of wind onshore installed capacities differences ({\si{\mega\watt}}) across all NUTS 2 regions (left) and mean Pearson correlation of wind onshore profiles across intersecting NUTS 2 regions (right).}
\label{fig_windonshore_inst_cap_Pearson_matrices_nuts2}
\end{figure}

\clearpage

\begin{figure}[htbp]
\centering
\includegraphics[width=0.75\linewidth]{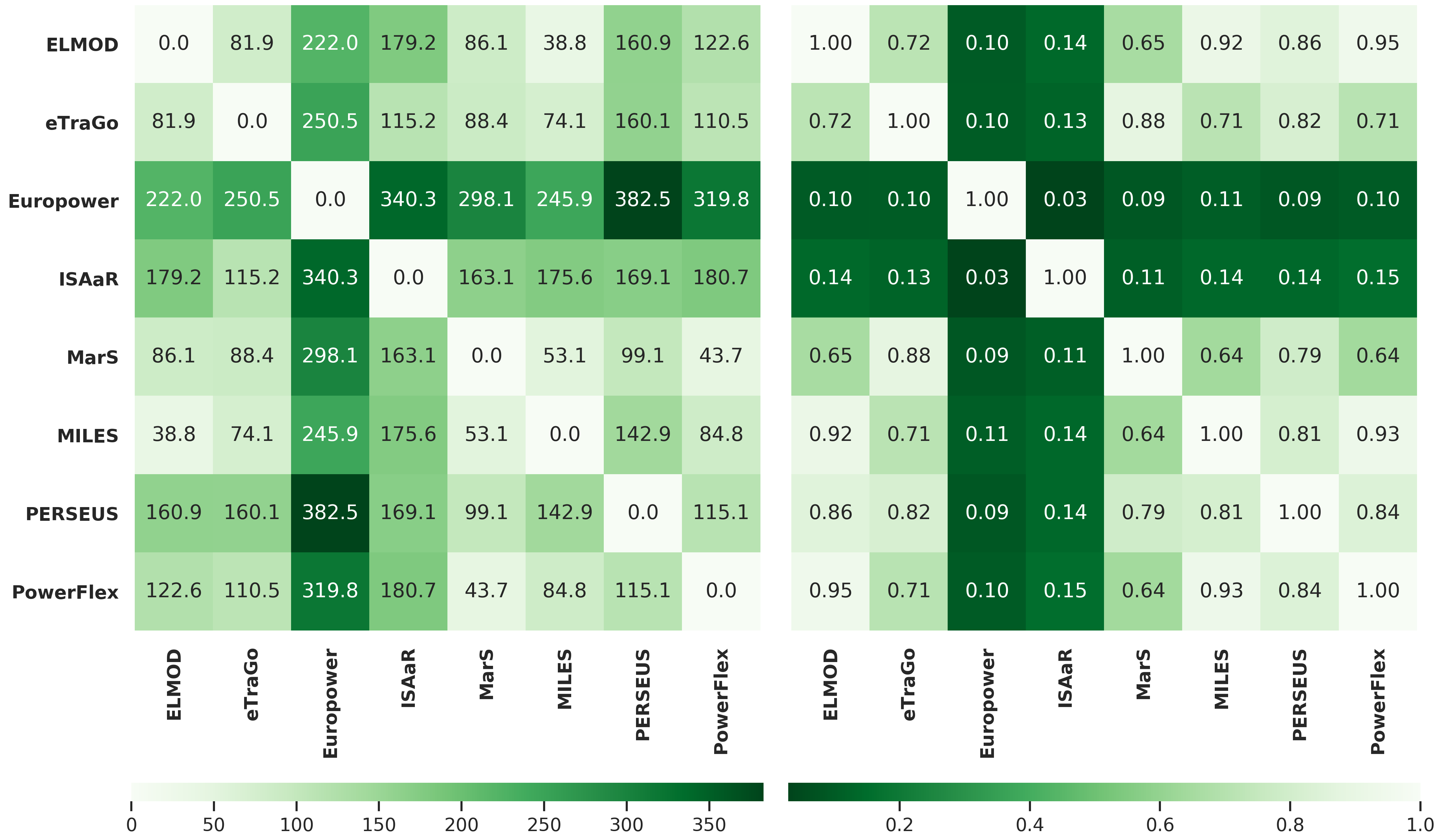}
\caption{\acs{RMSE} of wind offshore installed capacities differences ({\si{\mega\watt}}) across all NUTS 2 regions (left) and mean Pearson correlation of wind offshore profiles across intersecting NUTS 2 regions (right).}
\label{fig_windoffshore_inst_cap_Pearson_matrices_nuts2}
\end{figure}

\begin{figure}[htbp]
\centering
\includegraphics[width=0.75\linewidth]{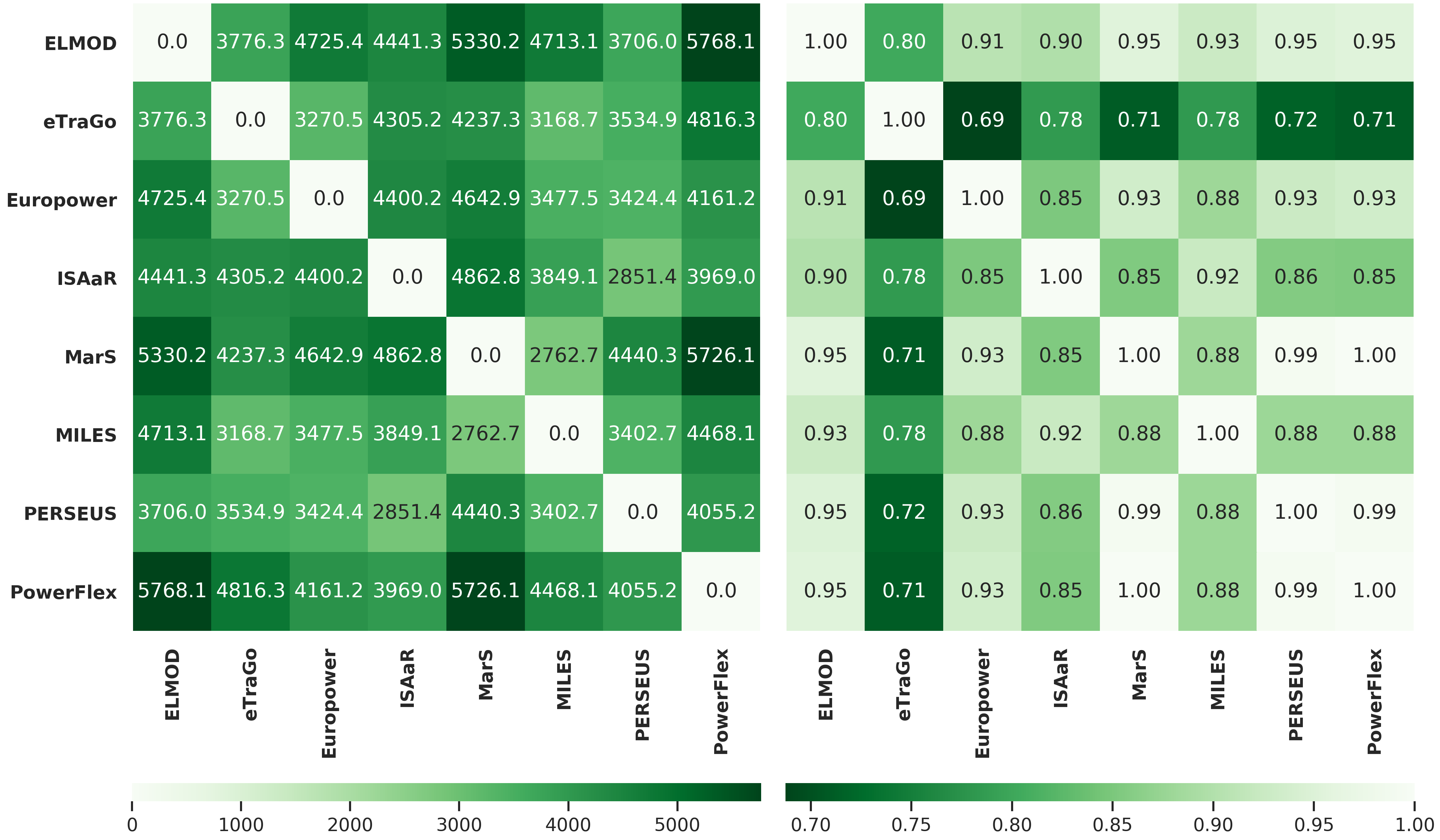}
\caption{\acs{RMSE} of annual load ({\si{\giga\watthour}}) differences across all NUTS 2 regions (left) and mean Pearson correlation of load profiles across intersecting NUTS 2 regions (right).}
\label{fig_load_inst_cap_Pearson_matrices_nuts2}
\end{figure}

\end{document}